\documentclass[aps,preprint,showpacs,superscriptaddress,groupedaddress]{revtex4-1}  
\pdfoutput=1
\usepackage{graphicx}  
\usepackage{dcolumn}   
\usepackage{bm}        
\usepackage{amssymb}   
\usepackage{amsmath}   
\usepackage{listings}
\usepackage{array}
\usepackage{graphicx}
\usepackage{lineno}
\usepackage{dcolumn}
\usepackage{color}
\usepackage{overpic}
\usepackage{multirow}
\usepackage{hyperref}

\newcommand{\bV}{\bold{V}}

\newcommand{\pT}{p_{\text{T}}}

\newcommand{\kp}{k^\prime}
\newcommand{\lp}{l^\prime}
\newcommand{\xmin}{x_{\min}}

\newcommand{\Nmodes}{N_{\text{modes}}}
\newcommand{\seff}{s_{\text{eff}}}
\newcommand{\bkg}{\text{bkg}}

\begin{document}

\widetext


\title{F2F, a model-independent method to determine the mass and width of a particle in the presence of interference}
\author{Li-Gang Xia \\ Department of Physics, Warwick University, CV4 7AL, UK}

\begin{abstract}
    It is generally believed that any particle to be discovered will have a TeV-order mass. Given its great mass, it must have a large decay width. Therefore, the interference effect will be very common if they and the Standard-Model (SM) particles contribute to the same final state. 
    However, the interference effect could make a new particle show up not like a resonance, and it is difficult to search and measure its properties.
    In this work, a model-independent method, F2F (Fit To Fourier coefficients), is proposed to search for an unknown resonance and to determine its mass ($M$) and width ($\Gamma$) in the presence of interference.
    Basically we express the sum of reosnant signal and the interference as a cosine Fourier series and relate the Fourier coefficients with the mass and width.
    The relation is based on the general propagator form, $1/(x^2-M^2+iM\Gamma)$. Thus it does not need any signal model. Toy experiments show that the obtained mass and width agree well with the inputs with a similar precision as using an explicit signal model. We also show that we can apply this method to measure the Stardard-Model Higgs width and to make statistic interpretation in searching for new resonance allowing for interference. 
\end{abstract}
\maketitle

\section{Introduction}
Many Beyond-Standard-Model (BSM) models predict heavy particles of the order of TeV. If we search with a final state to which the Standard-Model (SM) processes also contribute, some of the SM backgrounds will interfere coherently with the new signal. However, the interference effect could make things complicated. The new particle may appear as a bump, a dip or other strange structures (for experimental observation, see Ref.~\cite{phipi0}; for experimental seaches, see Ref.~\cite{ATLAS_hzz,ATLAS_vlq, CMS_zz, CMS_ttbar}; for phenominologic analyses, see Ref.~\cite{ggzz,ttbar,song}) in the distribution of an observable. 
Therefore it is difficult to search for and measure the properties of the new particle if we have not yet had enough knowledge about the interaction mechanism. This is particularly difficult for the searches at hardon colliders like the Large Hadron Collider (LHC). 

Let us consider a simplified inteference model.
Suppose $x$ is an observable (like the invariance mass of the final particles), the differential cross section is
\begin{equation}\label{eq:no1}
    \frac{d\sigma}{dx} \propto |A_{b}(x) + \frac{k(x)e^{i\delta(x)}}{x^2-M^2+iM\Gamma}|^2 
\end{equation}
Here $A_{b}$ is the background amplitude and the other term, a Breit-Wigner (BW) resonance signal, describes a new particle with mass $M$ and decay width $\Gamma$. $k(x)$ and $\delta(x)$ represent the strength and relative phase of the new amplitude. They could be functions of the observable. Without a good knowledge of the interaction mechanism, we cannot parameterize the inteference effect well and thus cannot make statistic interpretation easily 
(in other words, we have to make many simulations and interpretate under many hypotheses) and measure the mass and width by fitting. 

In this work, a novel method is proposed based on the fact that the squared propagator, $\frac{1}{(m^2-M^2)^2+M^2\Gamma^2}$, is the key factor for a resonance. The principle will be elaborated in next section. Many toy experiments are presented in Sec.~\ref{sec:toys} for different interference details and different detector responses. Application of this method to the SM Higgs width measurement and to search for new resonance allowing for interference will be shown in Sec.~\ref{sec:higgswidth} and Sec.~\ref{sec:newres}, respectively. We will summarize in Sec.~\ref{sec:summary}. 

\section{Principle of the method}\label{sec:principle}
The differential cross section in Eq.~\ref{eq:no1} can be expanded to be a sum of three terms. 
\begin{equation}\label{eq:no2}
    |A_{b}|^2 + \frac{k}{(x^2-M^2)^2+M^2\Gamma^2} + \frac{\sqrt{A_{b}k}((x^2-M^2)\cos\delta+M\Gamma\sin\delta)}{(x^2-M^2)^2+M^2\Gamma^2} 
\end{equation}
The first term is the background process; the second term is the signal resonance and the third term is their interference. Let $f(x)/((x^2-M^2)^2+M^2\Gamma^2)$ denote the combination of the latter two terms. We will call it ``effective siganl'' throughout this paper. Assuming $f(x)$ is a slow-varying function, the fast-varying component is mainly due to the factor $1/((x^2-M^2)^2+M^2\Gamma^2)$, namely, the mass and width. To measure the variation frequency, we can perform the Fourier transformation. 

Supposing we have $n$ measurements for an observable with the values $x_1,x_2,\cdots, x_n$, the probability distribution function (normalized to the number of entries), $p(x)$, could be written either using a sum of Dirac delta functions or as a Fourier series.
\begin{eqnarray}
    p(x) = &&\sum_{i=1}^n \delta(x-x_i) \:, \label{eq:delta} \\
    p(x) = &&\frac{c_0}{L}+ \sum_{k=1}^N c_k\cos\frac{k\pi (x-x_{\min})}{L} \label{eq:CFT}
\end{eqnarray}
where $L\equiv x_{\max}-x_{\min}$ and $x_{\min/\max}$ is the smallest/greatest value. 
Now let us investigate how $c_k$s depend upon $M$ and $\Gamma$ through the squared propagator. Defining $\kp\equiv k\pi/L$ for any $k$ and ignoring the slow-varying function $f(x)$, we have
\begin{equation}
    c_k \propto \Re\int_{-\infty}^{+\infty} \frac{e^{i\kp x}}{(x^2-M_0^2)^2+M_0^2\Gamma^2} dx \:,
\end{equation}
where $M_0\equiv M-x_{\min}$ (because the distribution is shifted by $x_{\min}$ as shown in Eq.~\ref{eq:CFT} so that it starts from 0 and ends at $L$). The integration can be done using the usual contour integration technique~\cite{loopint} and the result is
\begin{equation}
    c_k \propto \frac{\pi}{\gamma_0^{3/2}\sin\theta}e^{-|\kp|\sqrt{\gamma_0}\sin\frac{\theta}{2}}\cos(|\kp|\sqrt{\gamma_0}\cos\frac{\theta}{2}-\frac{\theta}{2}) \: .
\end{equation}
Here $\gamma_0 \equiv \sqrt{M_0^2(M_0^2+\Gamma^2)}$ and $\tan\theta\equiv \Gamma/M_0$. We can see that $c_k$ as a function of $k$ has two features. One is the exponential decay (mainly driven by non-vanishing $\Gamma$)  and the other is the oscillation (mainly driven by non-vanishing $M$). 

Furthermore, it is very convenient to consider the detector responce if a function is expressed as a Fourier seires. If the detector resolution is $\sigma$, we have 
\begin{eqnarray}
    p(x) = && \sum_k c_k\int_{-\infty}^{+\infty} \cos(\kp (y-x_{\min}))\frac{1}{\sqrt{2\pi}\sigma}e^{-\frac{(x-y)^2}{2\sigma^2}} dy \nonumber\\
    = && \sum_k c_ke^{-\frac{1}{2}(\kp\sigma)^2}\cos(\kp(x-x_{\min})) \:. \label{eq:cksmear} 
\end{eqnarray}
The equation above shows that we need to add an extra exponential factor $e^{-\frac{1}{2}(\kp\sigma)^2}$ for each Fourier coefficient. As a bonus, it also proves that unfolding is very difficult if we want to obtain the true distribution from the smeared distribution. Given the the features above, we are able to perform a fit to data to obtain the information on $M$ and $\Gamma$. There are two fitting ways. We will call them ``setup1'' and ``setup2'' below. 

For fitting setup1, we calculate the coefficients first for data and background samples according to the transformation law and the delta function form in Eq.~\ref{eq:delta}.
\begin{eqnarray}
    c_0 =&& \int p(x)dx = \sum_{i=1}^n 1 = n \:, \label{eq:c0} \\
    c_k =&&\frac{2}{L}\int p(x) \cos\frac{k\pi(x-x_{\min})}{L}dx = \frac{2}{L}\sum_{i=1}^n \cos\frac{k\pi(x_i-x_{\min})}{L} \: . \label{eq:ck}
\end{eqnarray}
To perform the fit, we need to know the covariance matrix because different $c_k$s are correlated. Treating $x_i$ ($i=1,2,\ldots,n$) as $n$ indepdendent and identically-distributed (iid) random variables, $c_k$ can be also seen as a random variable. The covariance matrix can be obtained in the standard way, namely,
\begin{equation} \label{eq:dck}
    \bV_{kl} = \frac{4}{L^2}\frac{1}{n}\sum_{i}^n[\cos(\kp(x-\xmin))\cos(\lp(x-\xmin))-\mu_k\mu_l]\:, \mu_k\equiv\frac{1}{n}\sum_{i=1}^n\cos(\kp(x-\xmin)) \:.
\end{equation}
where $\bV$ denotes the covariance matrix and the element $\bV_{kl}$ denotes the covariance of $c_k$ and $c_l$. After obtaining the Fourier coefficients for the data and background, their subtraction gives the Fourier coefficients for the effective signal. To get the best-estimated mass and width, we define the optimization function, $\chi^2(M,\Gamma)$, in the following way.
\begin{equation}
    \chi^2(M,\Gamma) \equiv \sum_{k=1}^{\Nmodes}(c_k(M,\Gamma)-c_k(\seff))[\bV^{-1}(\seff)]_{kl}(c_l(M,\Gamma)-c_l(\seff)) \:.
\end{equation}
Here $\Nmodes$ is the number of modes considered in the fit. We should choose it to be greater than $L/\sigma$. $c_k(\seff)$ is the subtraction of the $k$-mode coefficient in data and that in background samples. $\bV(\seff)$ is the corresponding covariance matrix. To be exact, they are defined below. $\bV^{-1}$ is the inverse matrix of $\bV$. 
\begin{eqnarray}
    c_k(\seff) \equiv c_k(\text{Data}) - c_k(\text{Bkg.}) \:,
    \bV(\seff) \equiv \bV(\text{Data}) + \bV(\text{Bkg.}) \:.
\end{eqnarray}
Given the principle above, we will call the method F2F ( Fit To Fourier coefficients). 

Fitting setup1 is basically a $\chi^2$ fit, which may have some bias for low-statistics fits. In fact, we are able to construct a likelihood fit (setup2). This is our recommendation. The total probability distribution function (PDF) is defined as
\begin{equation}\label{eq:fullPDF}
    f(x | M,\Gamma) \equiv \frac{1}{N_{\bkg}+c_0}(N_{\bkg}f_{\bkg}(x) + \frac{c_0}{L} + \sum_{k=1}^{\Nmodes} c_k(M,\Gamma) \cos(\kp(x-\xmin)) \: ,
\end{equation}
Here $c_0$ can be understood as the effective signal yield (see Eq.~\ref{eq:c0}) and could be negative due to interference. $f_{\bkg}(x)$ is the normalized background PDF. The total PDF, $f(M,\Gamma)$, is already normalized due to the Fourier form. The optimization function (we still call it $\chi^2(M,\Gamma)$) is then
\begin{equation}\label{eq:chi2}
    \chi^2(M,\Gamma) \equiv -2\ln\Pi_{i=1}^nf(x_i|M,\Gamma) = -2\sum_{i=1}^n\ln f(x_i|M,\Gamma)  \: .
\end{equation}

In practise, we are using the following form for the Fourier coefficients. 
\begin{equation}\label{eq:fitfunc}
    c_k(M,\Gamma, A, B, C)= A(k)+ Be^{-\frac{1}{2}(\kp\sigma)^2-\kp\sqrt{\gamma_0}\sin\frac{\theta}{2}}\cos(\kp \sqrt{\gamma_0}\cos\frac{\theta}{2}-C) 
\end{equation}
Here $A(k)$ is a polynomial function about the mode number $k$ (zeroth or first order is enough for toy experiments in next sections) to account for any background mismodeling. $B$ and $C$ are two floating parameters. They are intrdouced to account for the slow-varying functions not considered in the Fourier transformation. 

In the end of this section, it is interesting to look closer at the exponential factor in Eq.~\ref{eq:fitfunc}. For a narrow-width resonance like the SM Higgs boson, we have $\Gamma<<M$ and this factor is approximately
\begin{equation}
    e^{-\frac{1}{2}\left(\frac{k\pi\sigma}{L}\right)^2 - \frac{1}{2}\frac{k\pi\Gamma}{L}}.
\end{equation}
If the two terms are comparable, we have
\begin{equation}
    \left(\frac{k\pi\sigma}{L}\right)^2 = \frac{k\pi\Gamma}{L} \: \Rightarrow \: \Gamma = \frac{k\pi\sigma^2}{L} \: .
\end{equation}
Therefore we can use $\frac{\sigma^2}{L}$ as a crude estimation of the ideal sensitivity on measuring the width $\Gamma$. For a mass range $L=1$~TeV and resolution $\sigma=1$~GeV, this method seems to be able to probe a width of the order of MeV. This is essentially what we need in the case of SM Higgs width measurement.

\section{Toy experiments}\label{sec:toys}
\begin{figure}
    \includegraphics[width=0.35\textwidth]{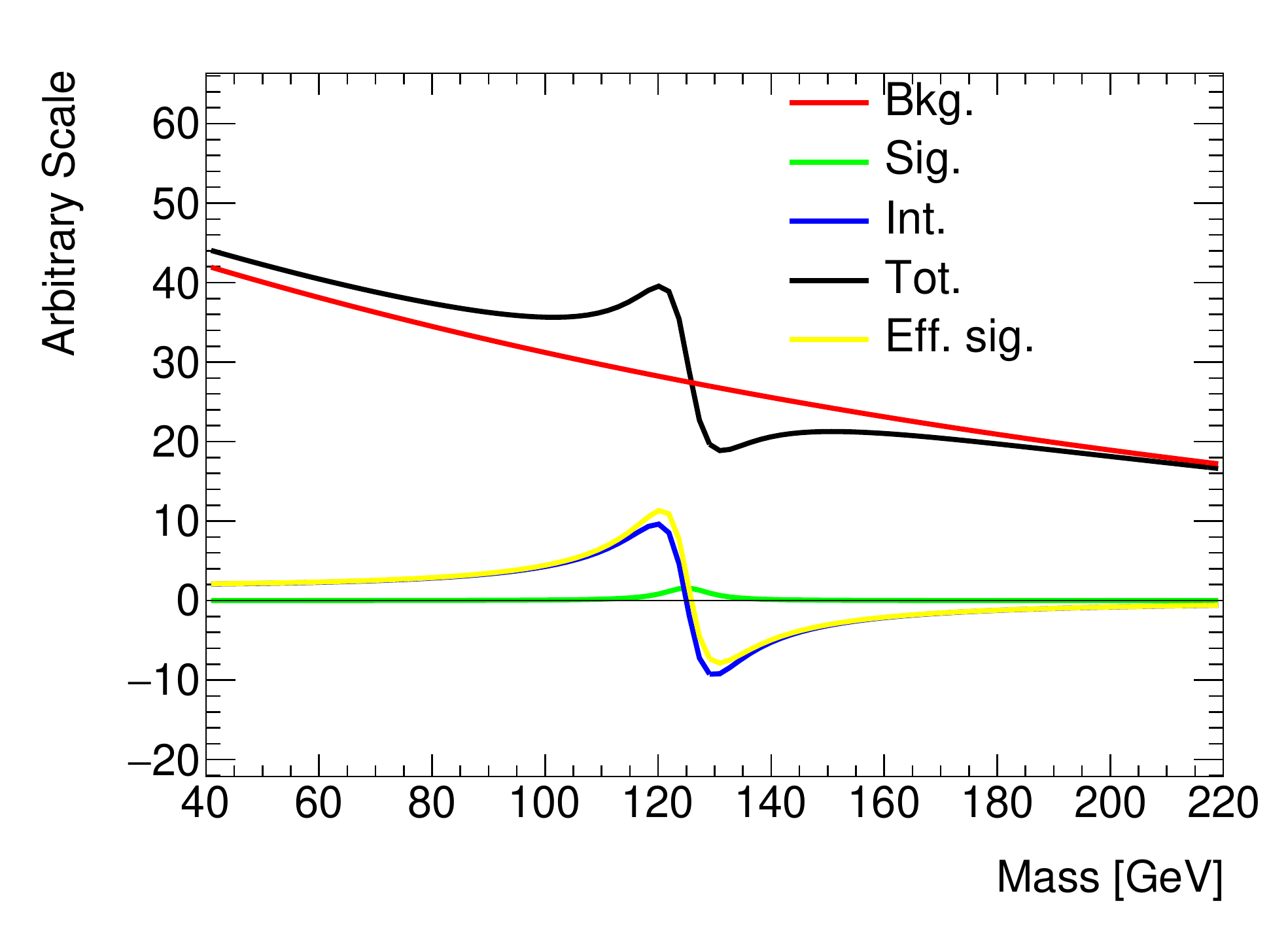}
    \includegraphics[width=0.35\textwidth]{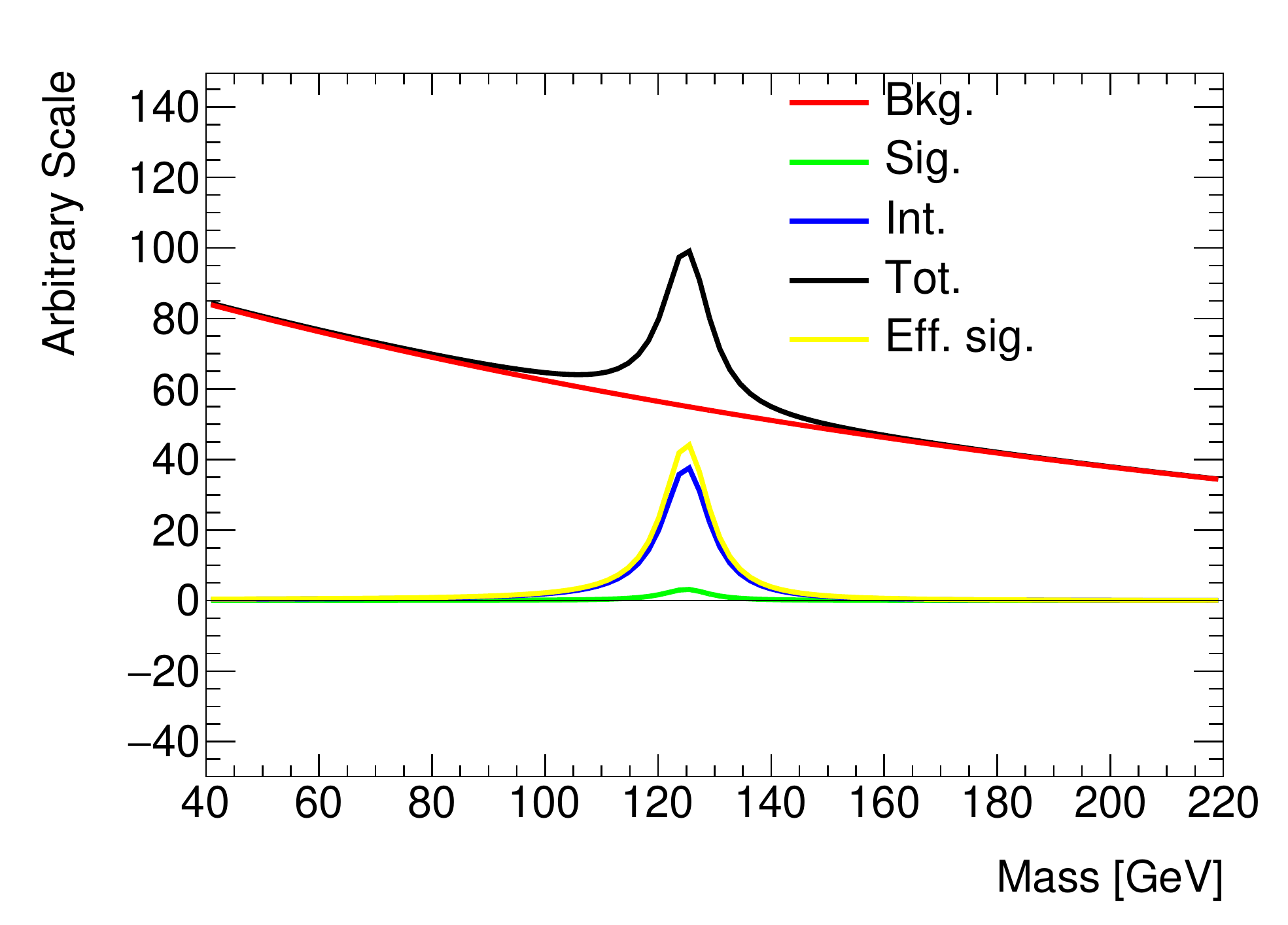}\\
    \includegraphics[width=0.35\textwidth]{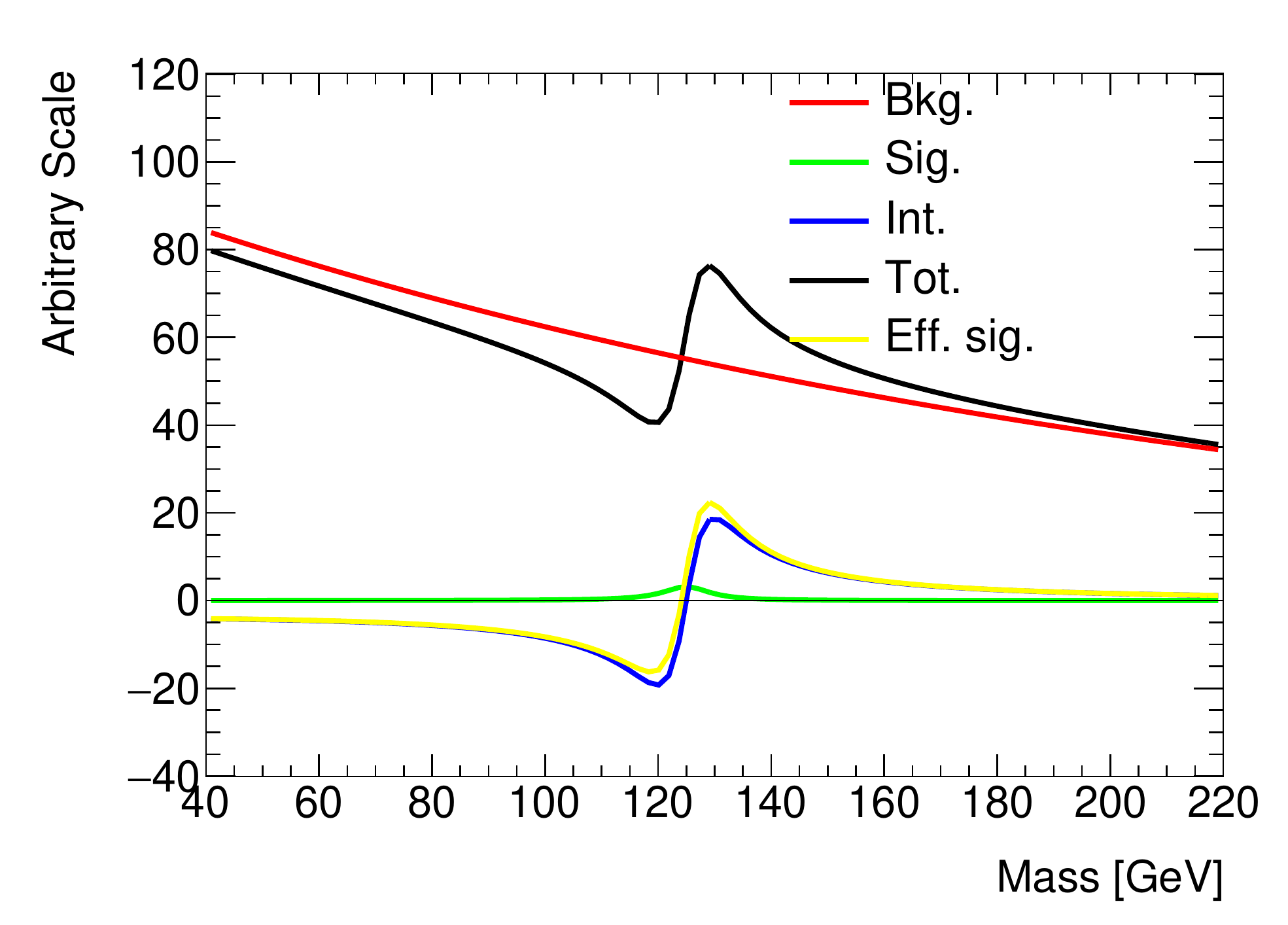}
    \includegraphics[width=0.35\textwidth]{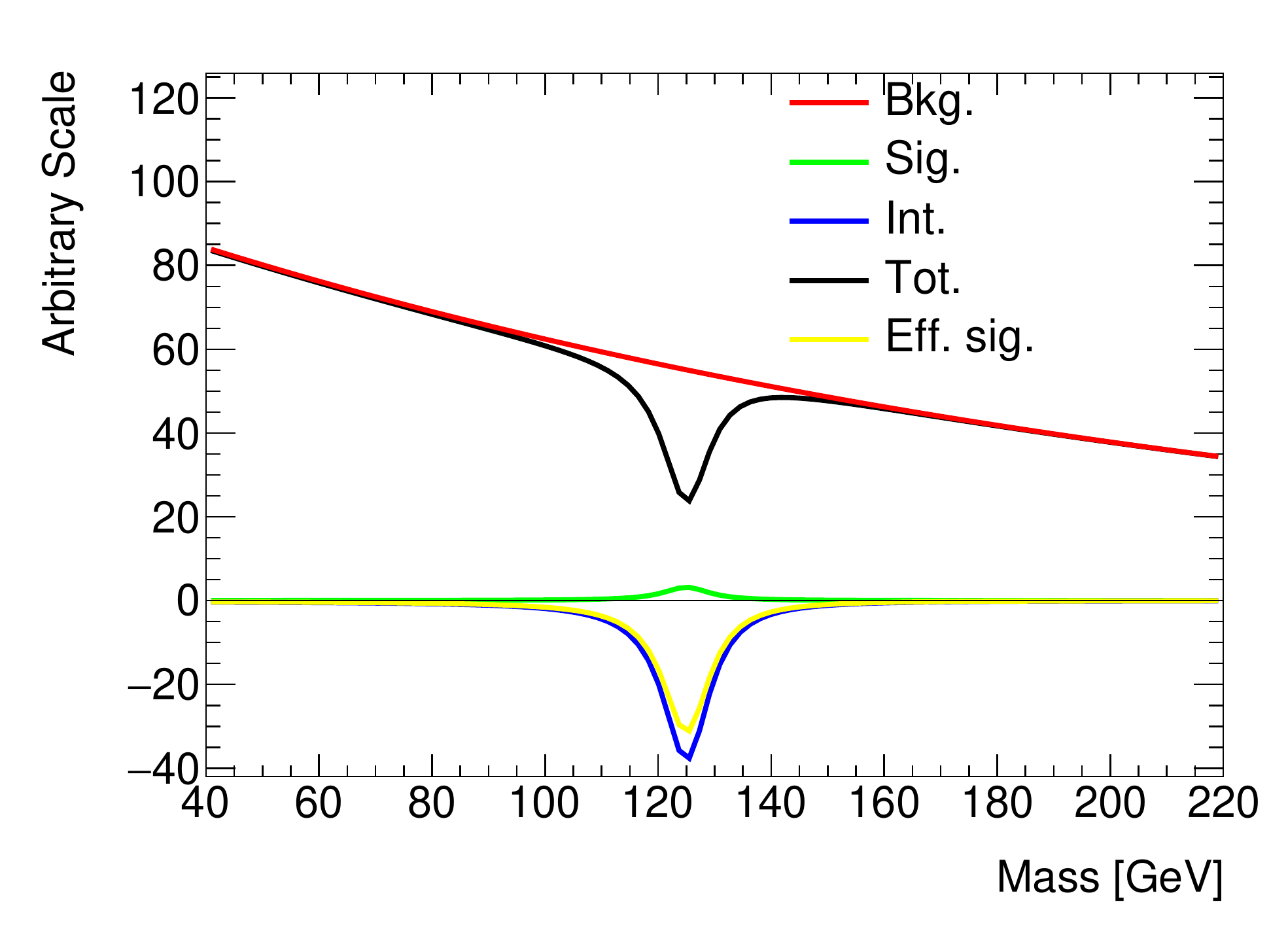}
    \caption{\label{fig:funcs}
        (color online) Differential cross sections for different phase angles, $\delta=180^\circ$ (top left), $\delta = 90^\circ$ (top right), $\delta = 0^\circ$ (bottom left) and $\delta = -90^\circ$ (bottom right). The red curve represents the background; green curve represents the signal; blue curve represents the interference contribution; black curve is the sum of all terms and the yellow curve represents the sum of signal and interference. 
    }
\end{figure}

In this section, let us investigate the performance of this method based on the following example. Suppose we have an exponential background and a Breit-Wigner resonance signal and they interfere coherently. The resonance has a mass, $M=125$~GeV, and full width, $\Gamma=10$~GeV. To be exact, the differential cross section with respective to a mass variable,$x$, is
\begin{equation}\label{eq:model}
    \frac{d\sigma}{dx}\propto |\frac{\sqrt{b}}{\sqrt{\tau}}e^{-\frac{x}{2\tau}} + \frac{\sqrt{kbr}e^{i\delta}}{x^2-M^2+iM\Gamma}|^2\:,
\end{equation}
where $b$ represents the background yield; $k=\frac{2\sqrt{2}M\Gamma\gamma}{\pi\sqrt{M^2+\gamma}}$ with $\gamma\equiv \sqrt{M^2(M^2+\Gamma^2)}$; $\tau=200$~GeV is the decaying length for the background distribution; and $r=0.005$ is the signal-to-background yield ratio. The mass distributions for 4 relative phases ($\delta = 180^\circ$, $90^\circ$ , $0^\circ$ and $-90^\circ$) are shown in Fig.~\ref{fig:funcs}. We can see very different structures for the same resonance state.

\subsection{Perfect detector resolution}
\begin{figure}
    \includegraphics[width=0.35\textwidth]{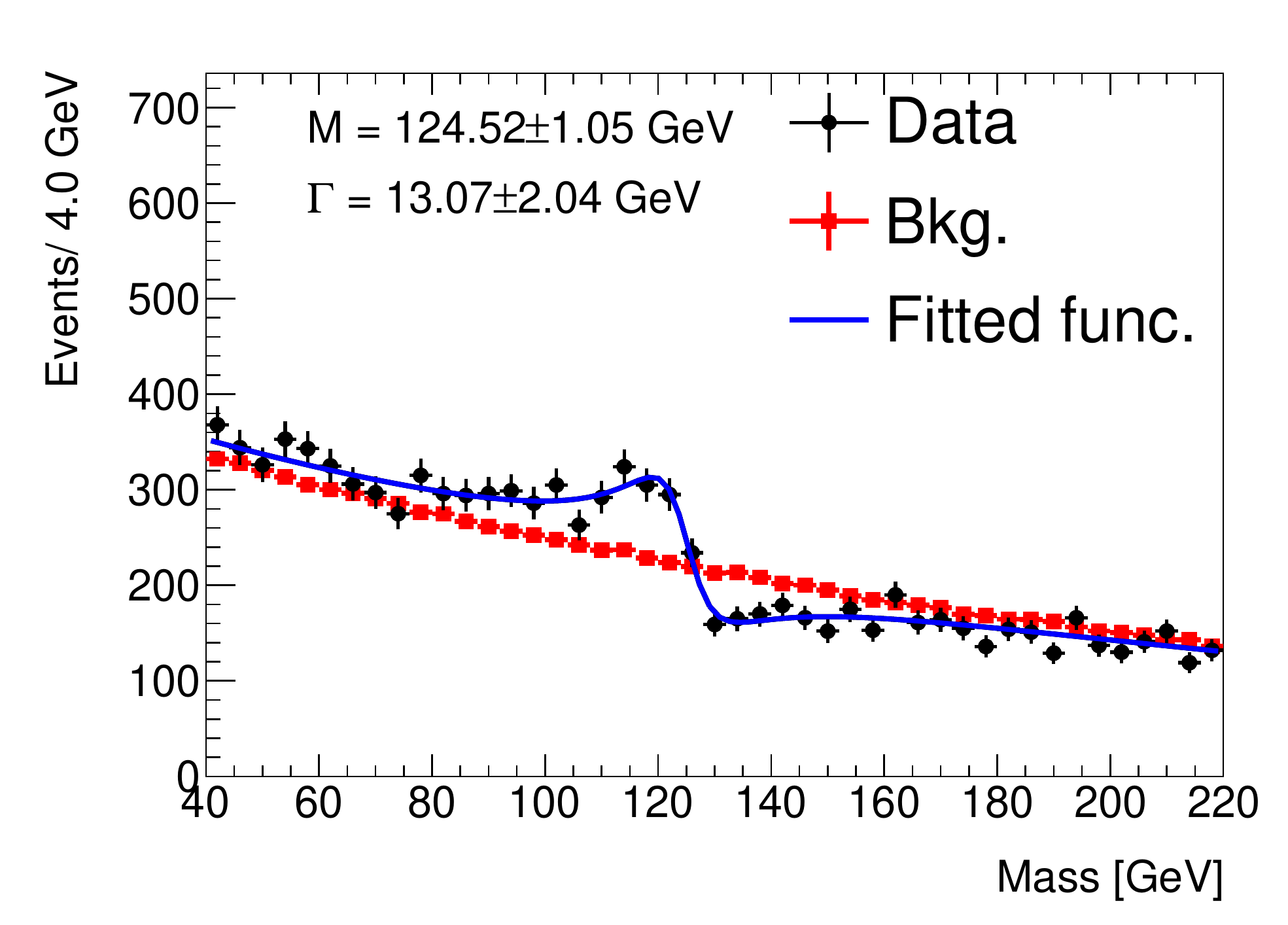}
    \includegraphics[width=0.35\textwidth]{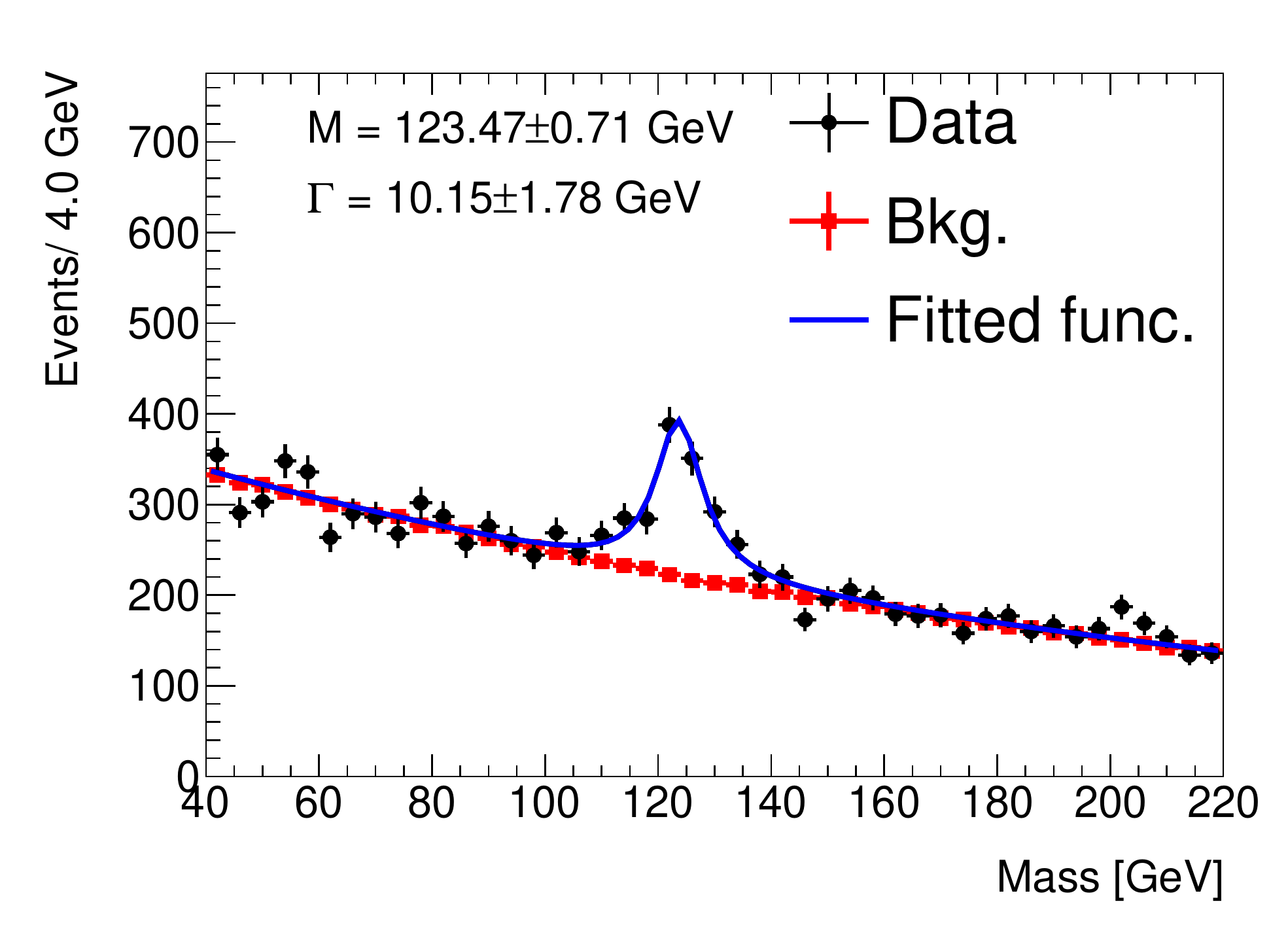}\\
    \includegraphics[width=0.35\textwidth]{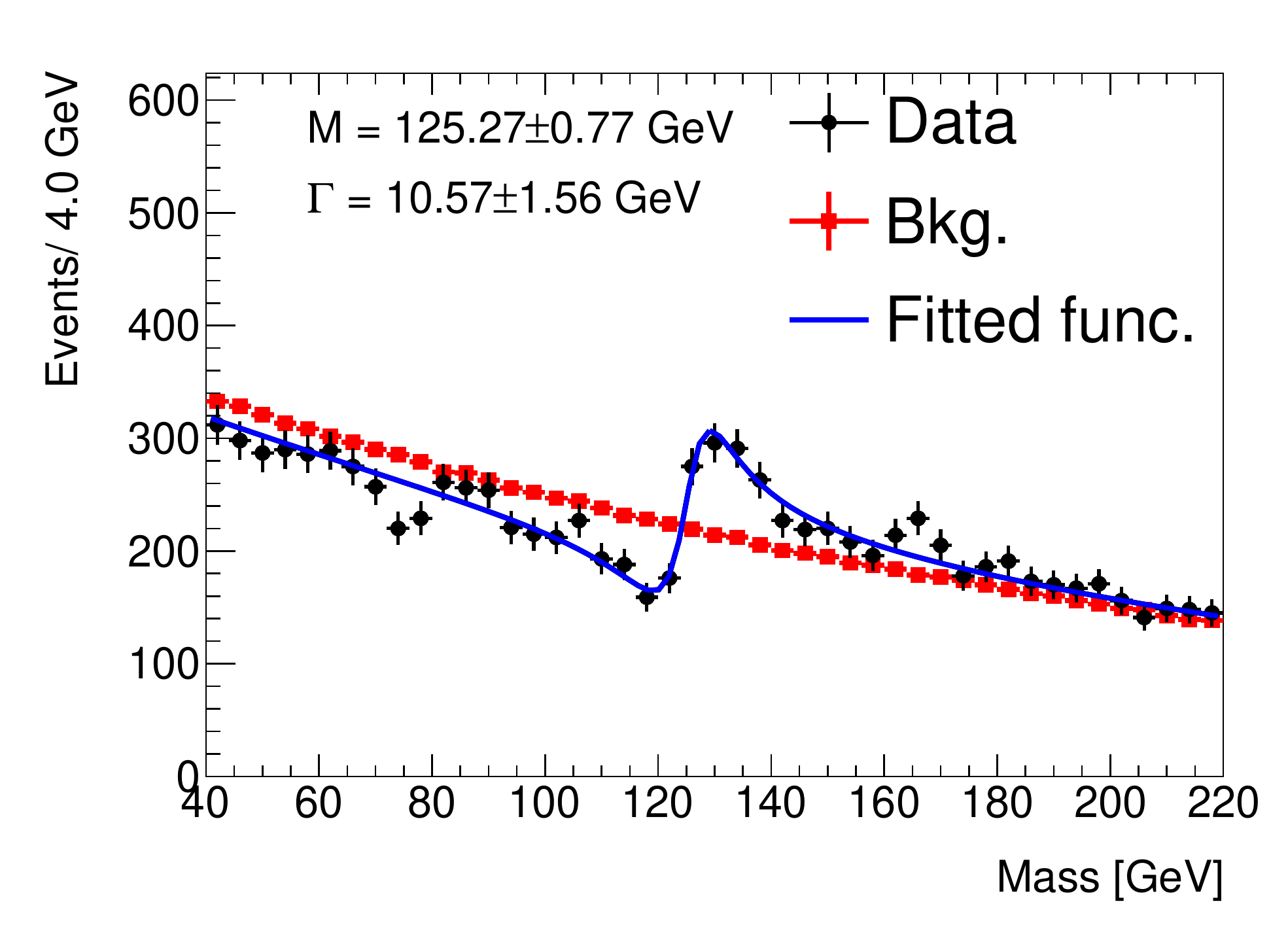}
    \includegraphics[width=0.35\textwidth]{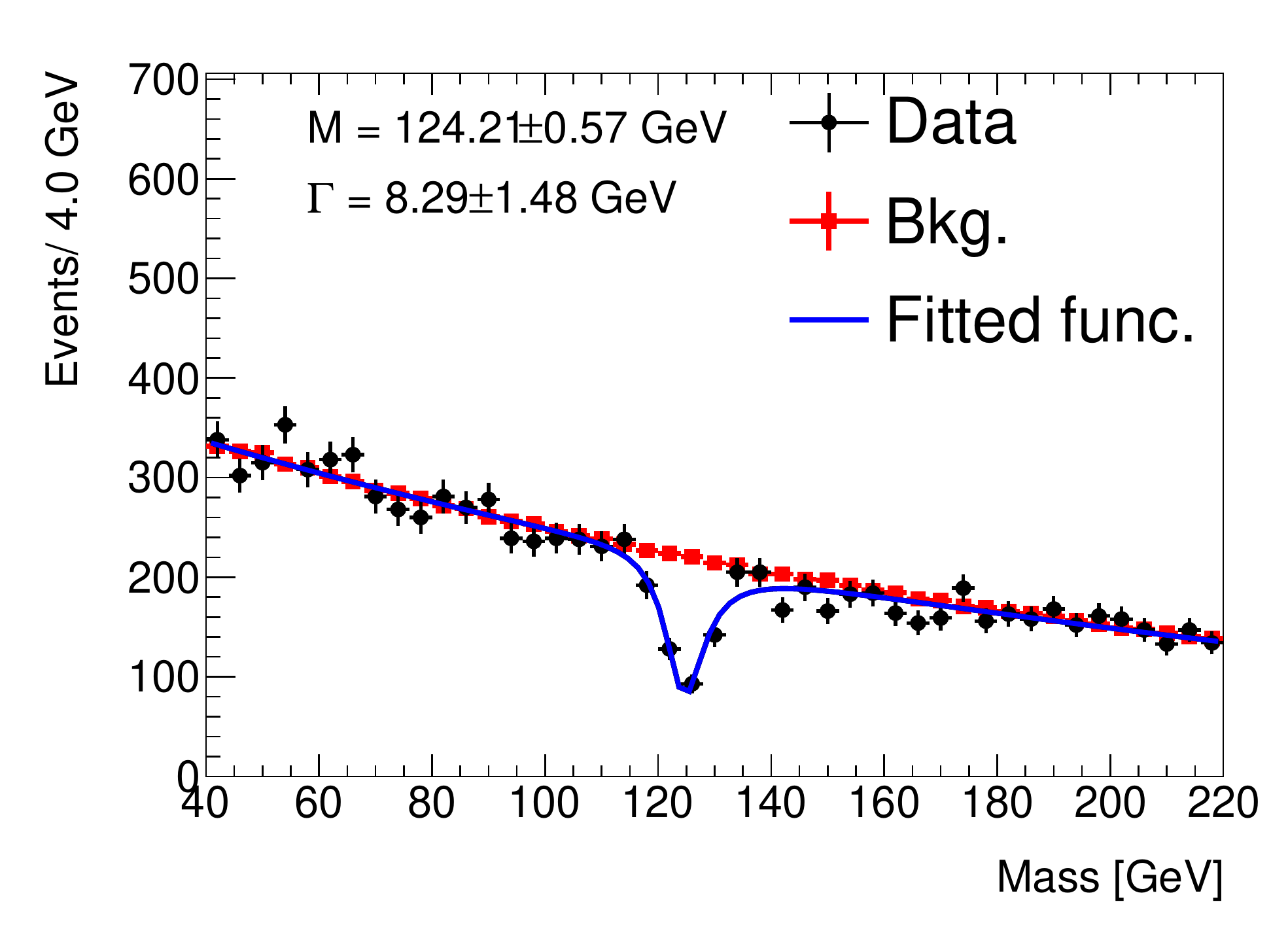}
    \caption{\label{fig:funcfit0}
        (color online) 
        Fitting results for different phase angles, $\delta=180^\circ$ (top left), $\delta = 90^\circ$ (top right), $\delta = 0^\circ$ (bottom left) and $\delta = -90^\circ$ (bottom right). The black dots with error bar represent the data sample. The red squares represent the background sample. The blue curves represent the fit results. 
    }
\end{figure}

In the first place, we do not consider the effect of the detector resolution. For each relative phase, a data sample, a background sample  and a signal sample (the method itself does not need the signal sample) are produced. Figure~\ref{fig:funcfit0} shows the fitting results using the explict signal model, namely, the formula in Eq.~\ref{eq:model}. They will be compared with the method proposed in this work. 

\begin{figure}
    \includegraphics[width=0.35\textwidth]{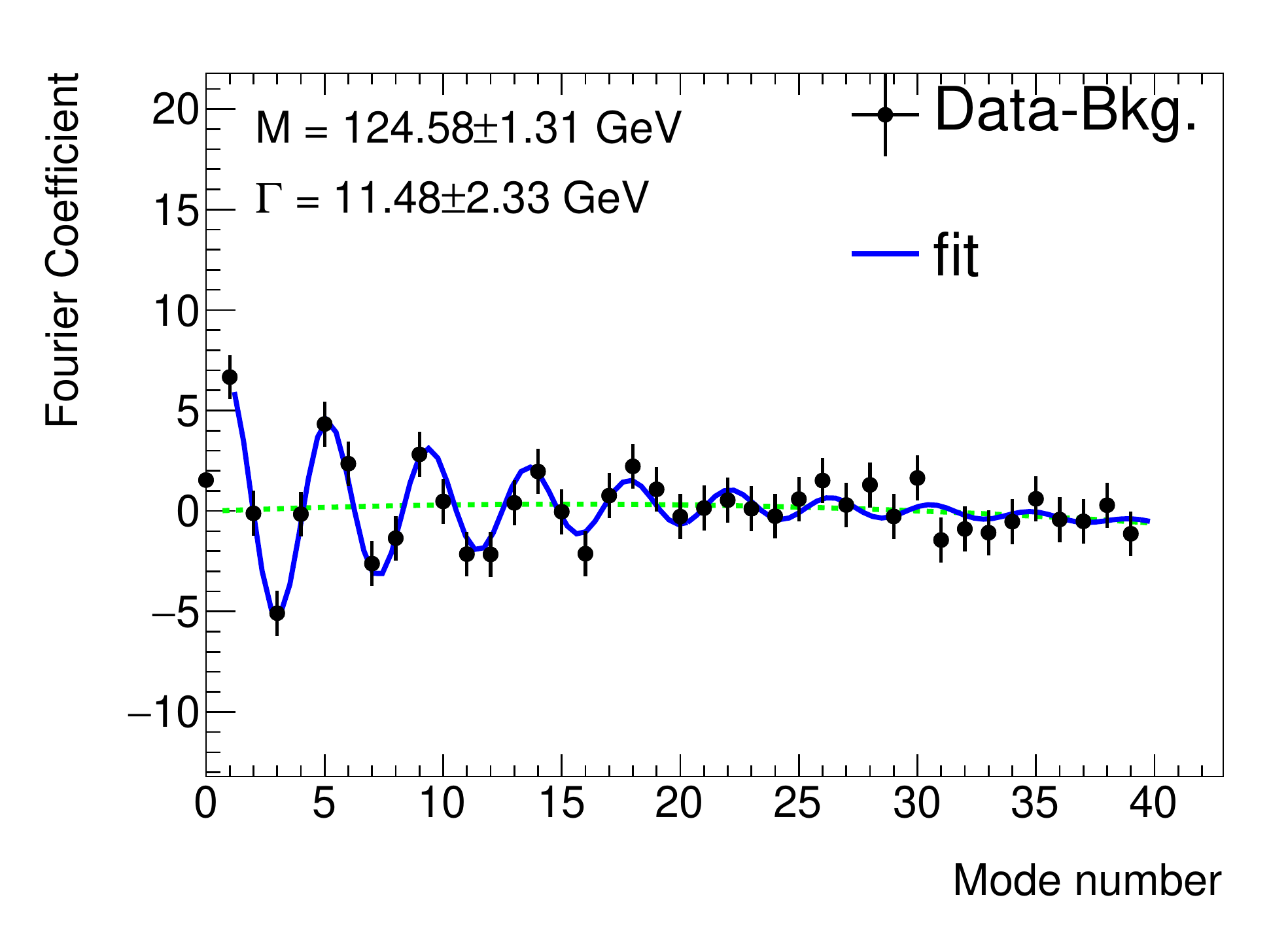}
    \includegraphics[width=0.35\textwidth]{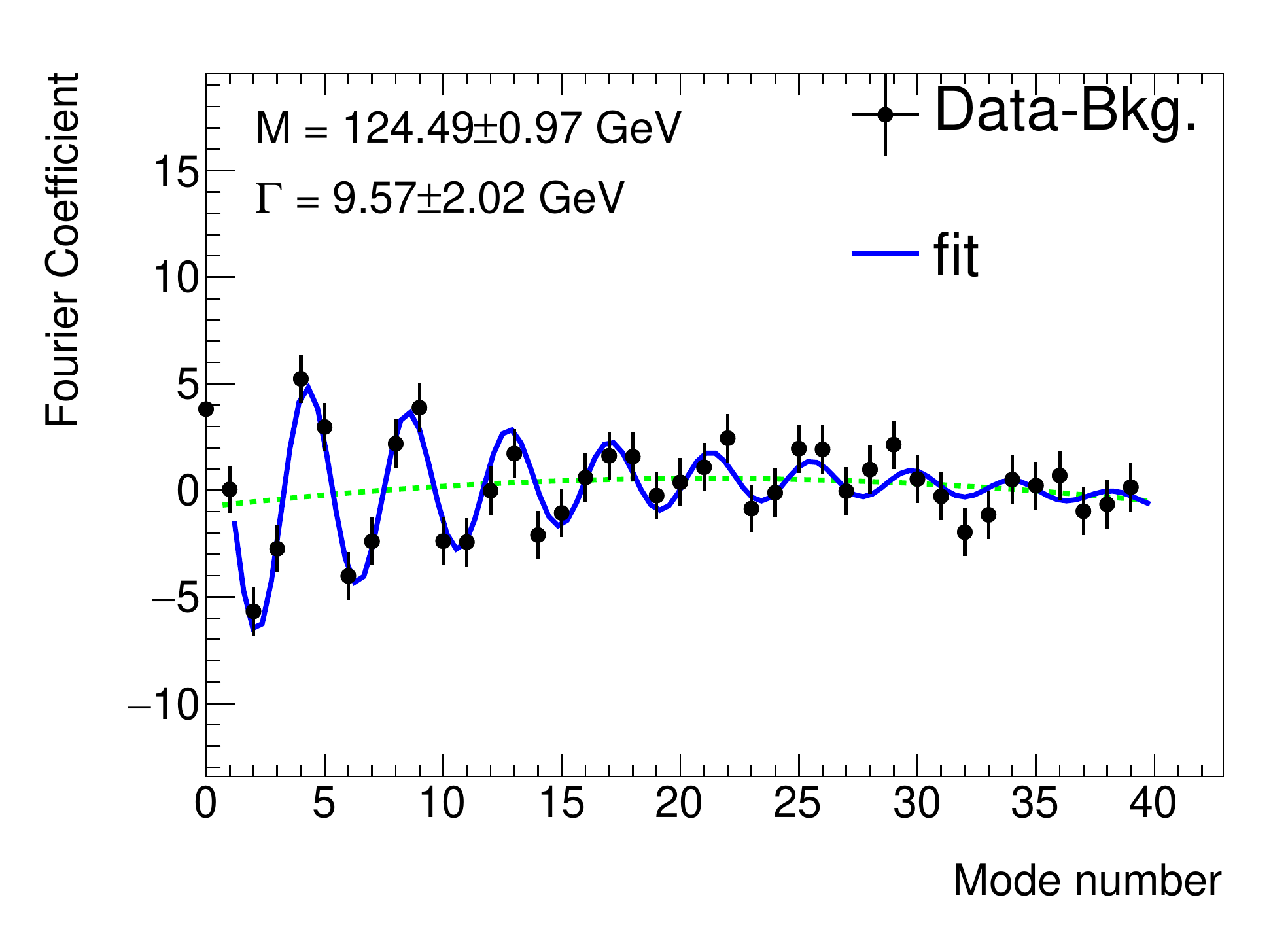}\\
    \includegraphics[width=0.35\textwidth]{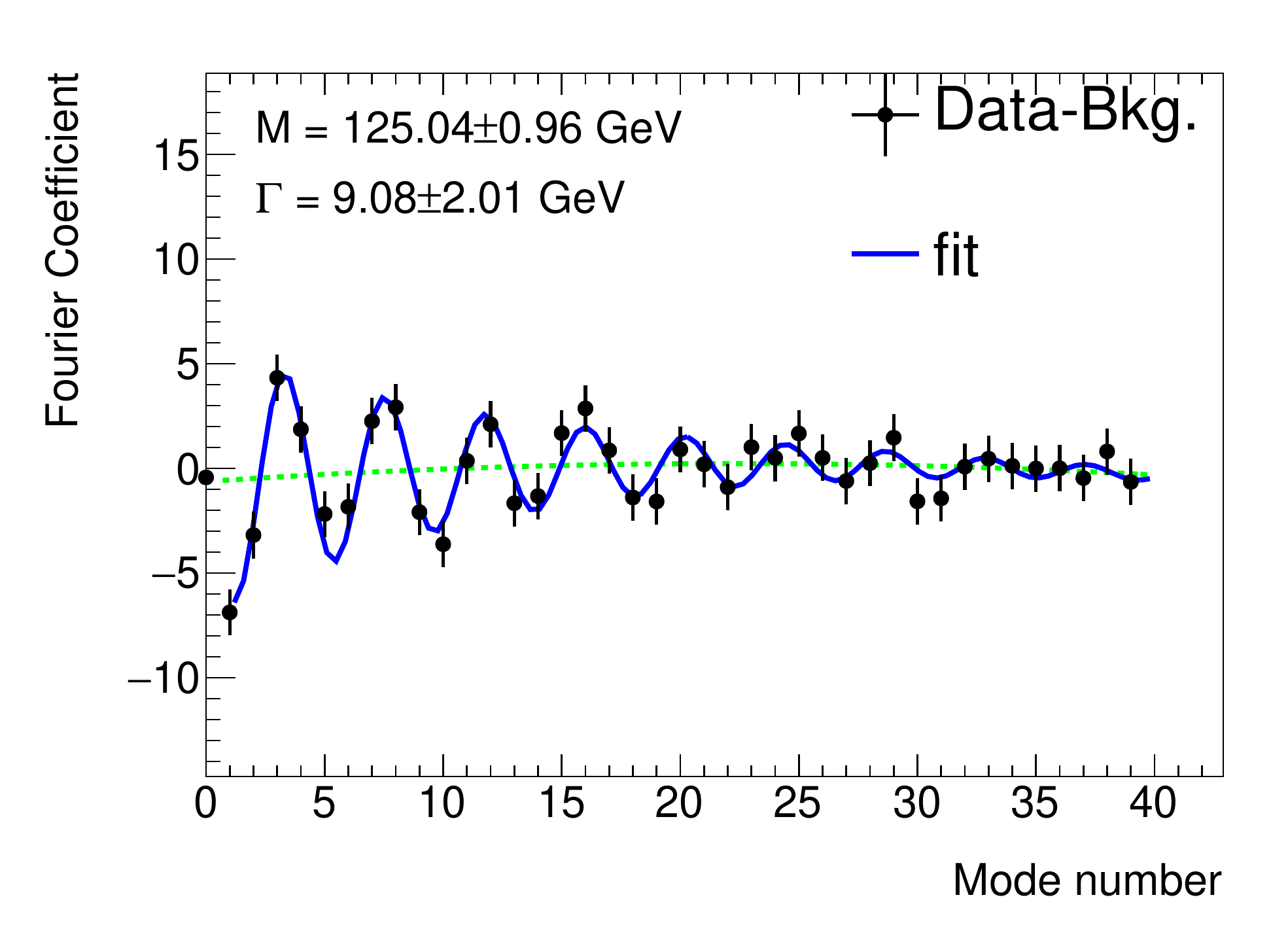}
    \includegraphics[width=0.35\textwidth]{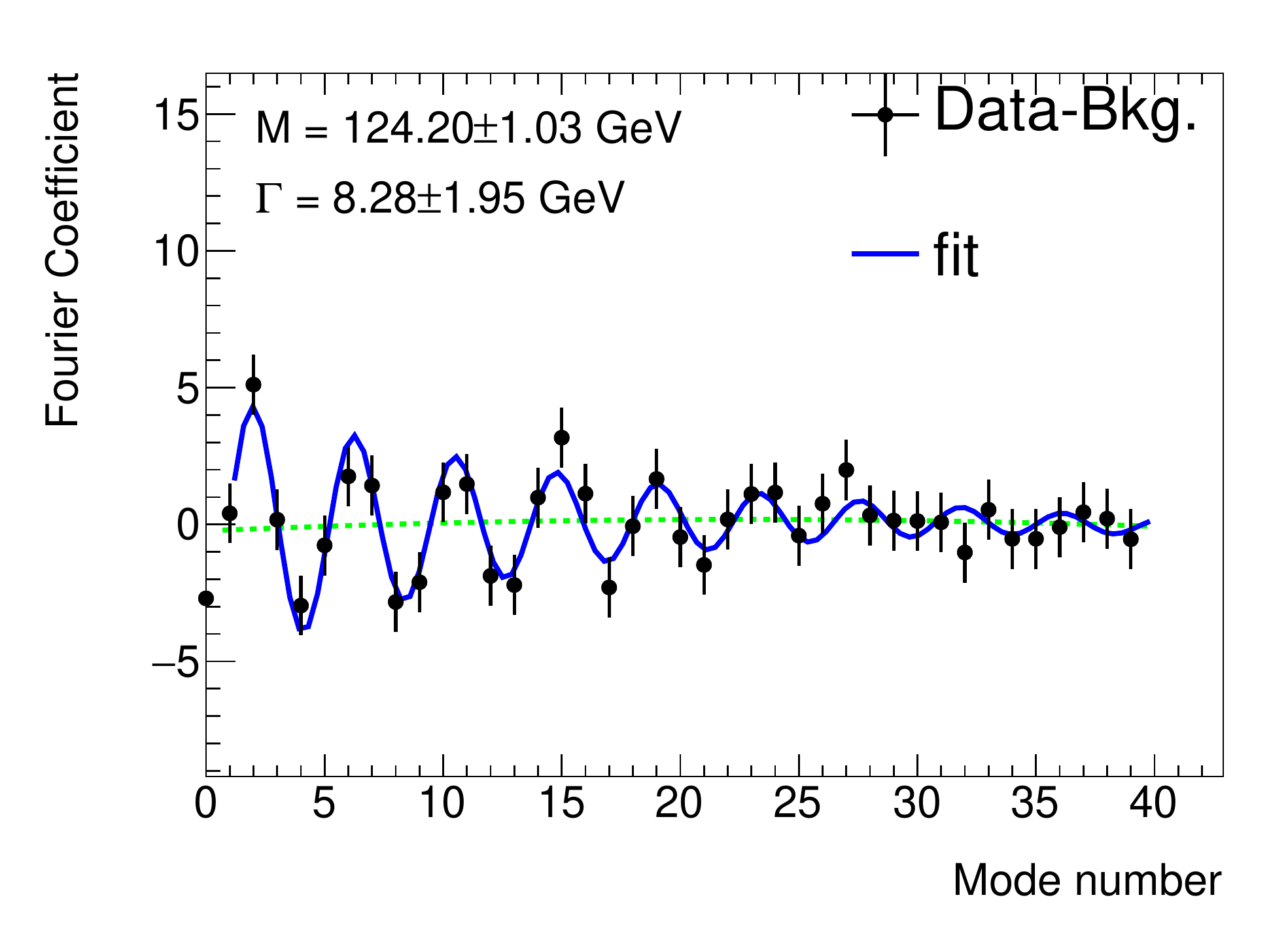}
    \caption{\label{fig:ckfitsigeff0}
        (color online) 
        Fitting results for different phase angles, $\delta=180^\circ$ (top left), $\delta = 90^\circ$ (top right), $\delta = 0^\circ$ (bottom left) and $\delta = -90^\circ$ (bottom right). The black dots with error bar represent the Fourier coefficients of the effective signal (data subtracting background). The blue curves represent the full fitting results. The green dashed curves represents the slow-varying contribution, namely, the polynominal part in Eq.~\ref{eq:fitfunc}. 
    }
\end{figure}

\begin{figure}
    \includegraphics[width=0.35\textwidth]{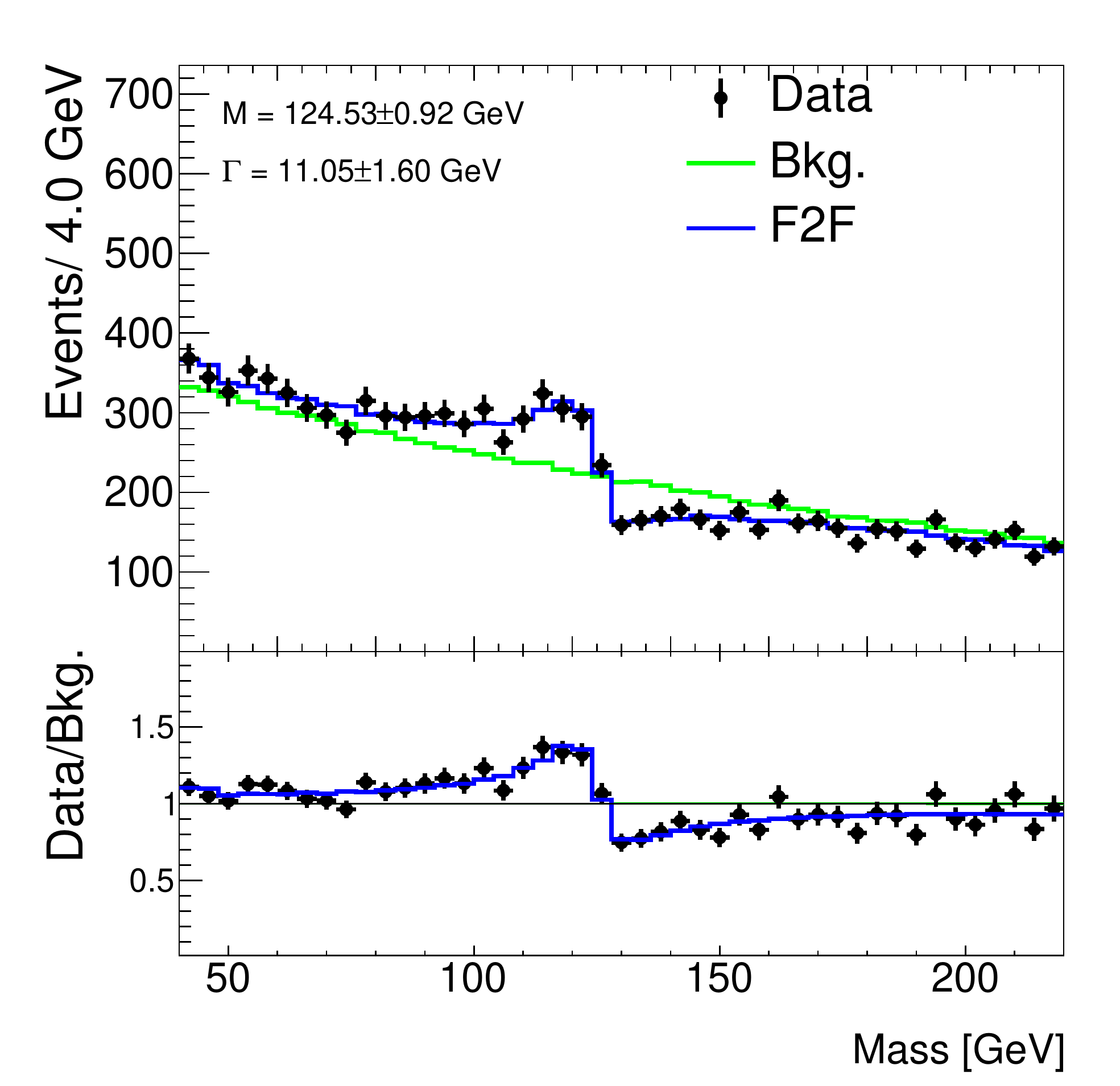}
    \includegraphics[width=0.35\textwidth]{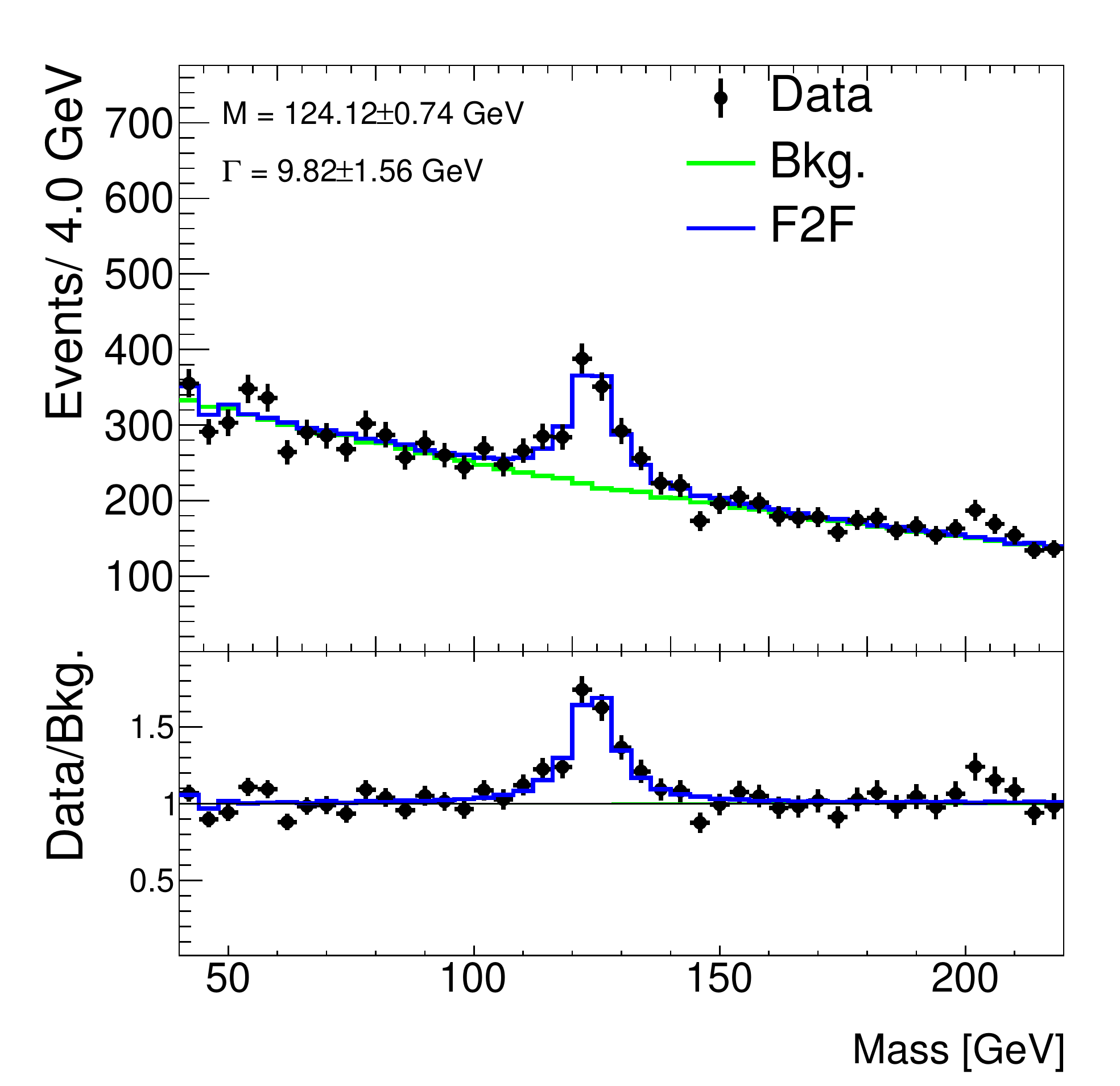}\\
    \includegraphics[width=0.35\textwidth]{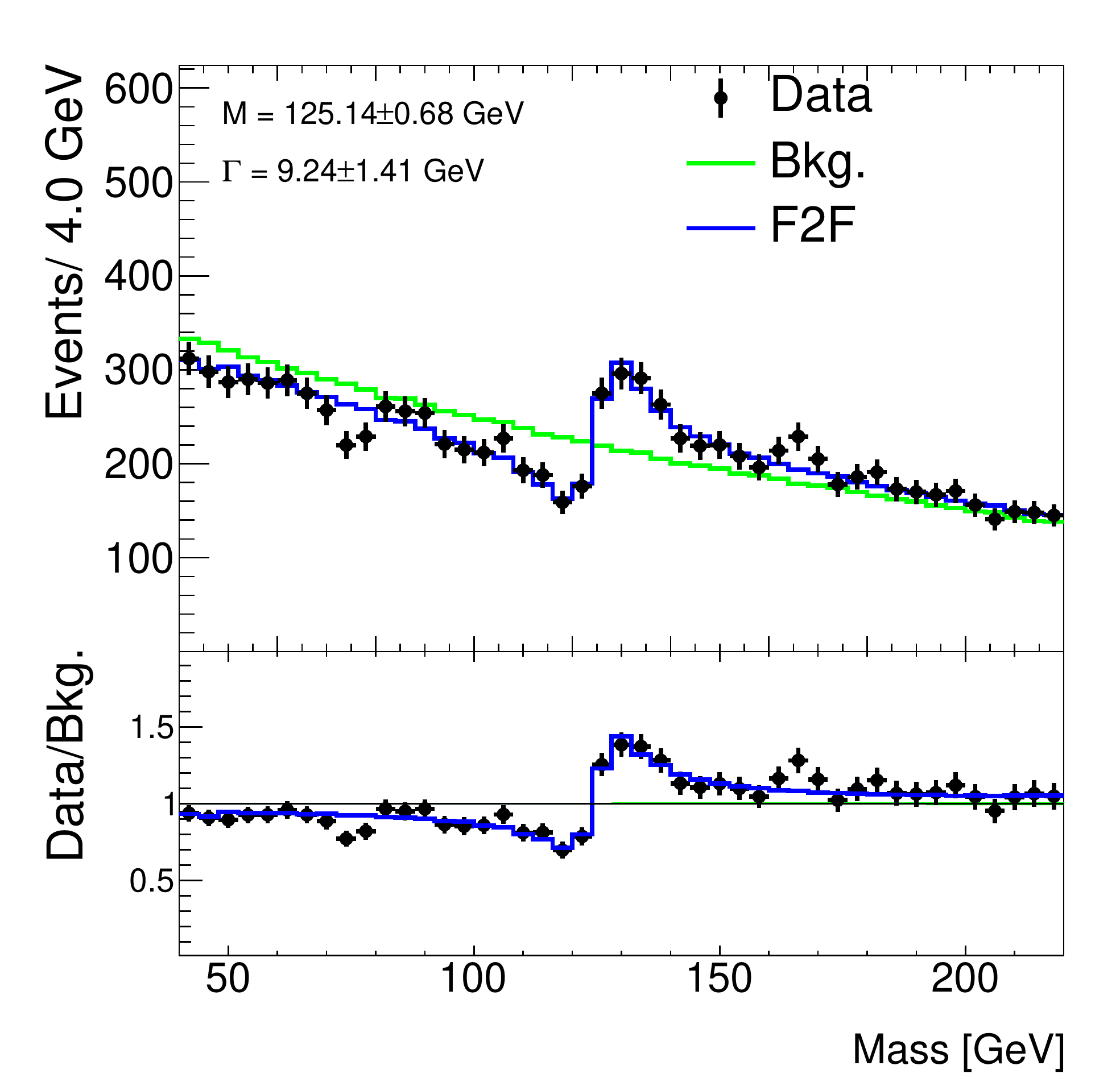}
    \includegraphics[width=0.35\textwidth]{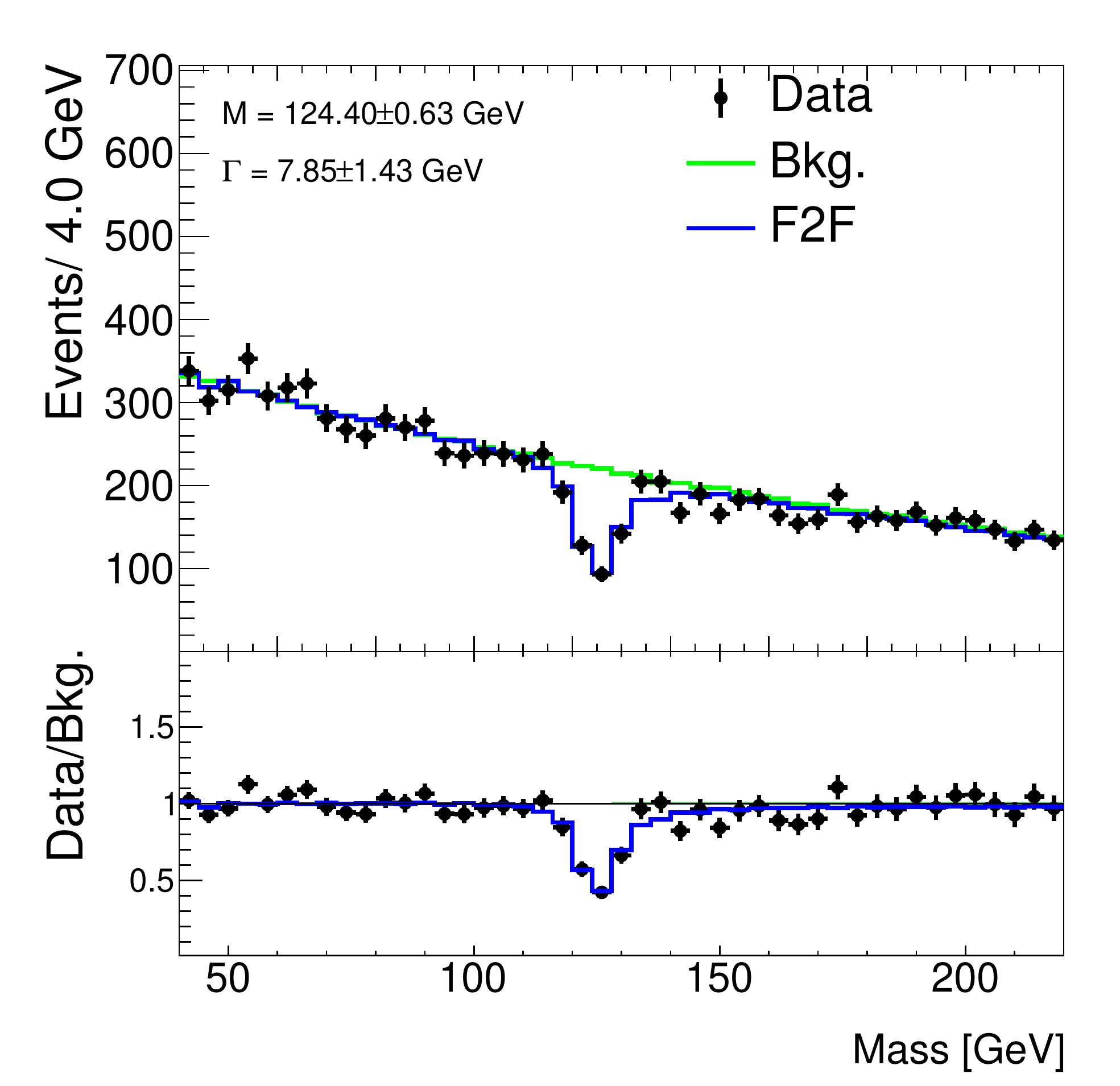}
    \caption{\label{fig:ckfitsigeff0p}
        (color online) 
        Fitting results for different phase angles, $\delta=180^\circ$ (top left), $\delta = 90^\circ$ (top right), $\delta = 0^\circ$ (bottom left) and $\delta = -90^\circ$ (bottom right). The black dots with error bar represent the Fourier coefficients of the effective signal (data subtracting background). The green histograms represent the background. The blue curves represent the full fitting results. 
    }
\end{figure}

To use the F2F method for fitting setup1, we calculate the Fourier coefficients for the data and background samples according to Eq.~\ref{eq:ck} and Eq.~\ref{eq:dck}. The Fourier coefficients for the effective signal are obtained by their subtraction. Fitting Eq.~\ref{eq:fitfunc} to these coefficients gives the final results. They are shown in Fig.~\ref{fig:ckfitsigeff0} and we can clearly see the two features, exponental decaying and oscillation about mode number. 
The fitting results for setup2 are shown in Fig.~\ref{fig:ckfitsigeff0p}. All results are summarized in Table~\ref{tab:fit0}. They agree well with each other and also agree with the inputs. The difference between setup1 and setup2 is probably due to the different fitting strategies. The setup2 is always better than setup1. Therefore we will use setup2 in next sections. The difference between the nominal method (i.e. using explicit signal model) and this method is probably because background
modeling is slightly different. The former uses the exponential form with
background yield floating while
the latter uses the monte-carlo (MC) prediction directly with both background yield and shape fixed.

\begin{table}
    \caption{\label{tab:fit0}
        Summary of the fitting results on the mass and width in the case of perfect detector resolution (unit: GeV) 
    }
    \begin{ruledtabular}
        \begin{tabular}{l | l l | l l | l l}
            & \multicolumn{2}{c}{Explicit signal model}  & \multicolumn{2}{c}{ F2F (setup1) } & \multicolumn{2}{c}{ F2F (setup2) }  \\
            \hline
            Phase& Mass  & Width  & Mass  & Width  & Mass  & Width \\
            \hline
            $180^\circ$ & $124.52\pm1.05$ & $13.07\pm2.04$  & $124.58\pm1.31$ & $11.48\pm2.33$ &$124.53\pm0.92$ & $11.05\pm1.60$ \\
            $90^\circ$  & $123.47\pm0.71$ & $10.15\pm1.78$  & $124.49\pm0.97$ & $9.57\pm2.02$  &$124.12\pm0.74$ & $9.82\pm1.56$ \\
            $0^\circ$   & $125.27\pm 0.77$ & $10.57\pm1.56$ & $125.04\pm0.96$ & $9.08\pm2.01$  &$125.14\pm0.68$ & $9.24\pm1.41$ \\
            $-90^\circ$ & $124.21\pm0.57$ & $8.29\pm1.48$   & $124.20\pm1.03$ & $8.28\pm1.95$  &$124.40\pm0.63$ & $7.85\pm1.43$ \\
        \end{tabular}
    \end{ruledtabular}
\end{table}

\subsection{Imperfect detector resolution}
\begin{figure}
    \includegraphics[width=0.32\textwidth]{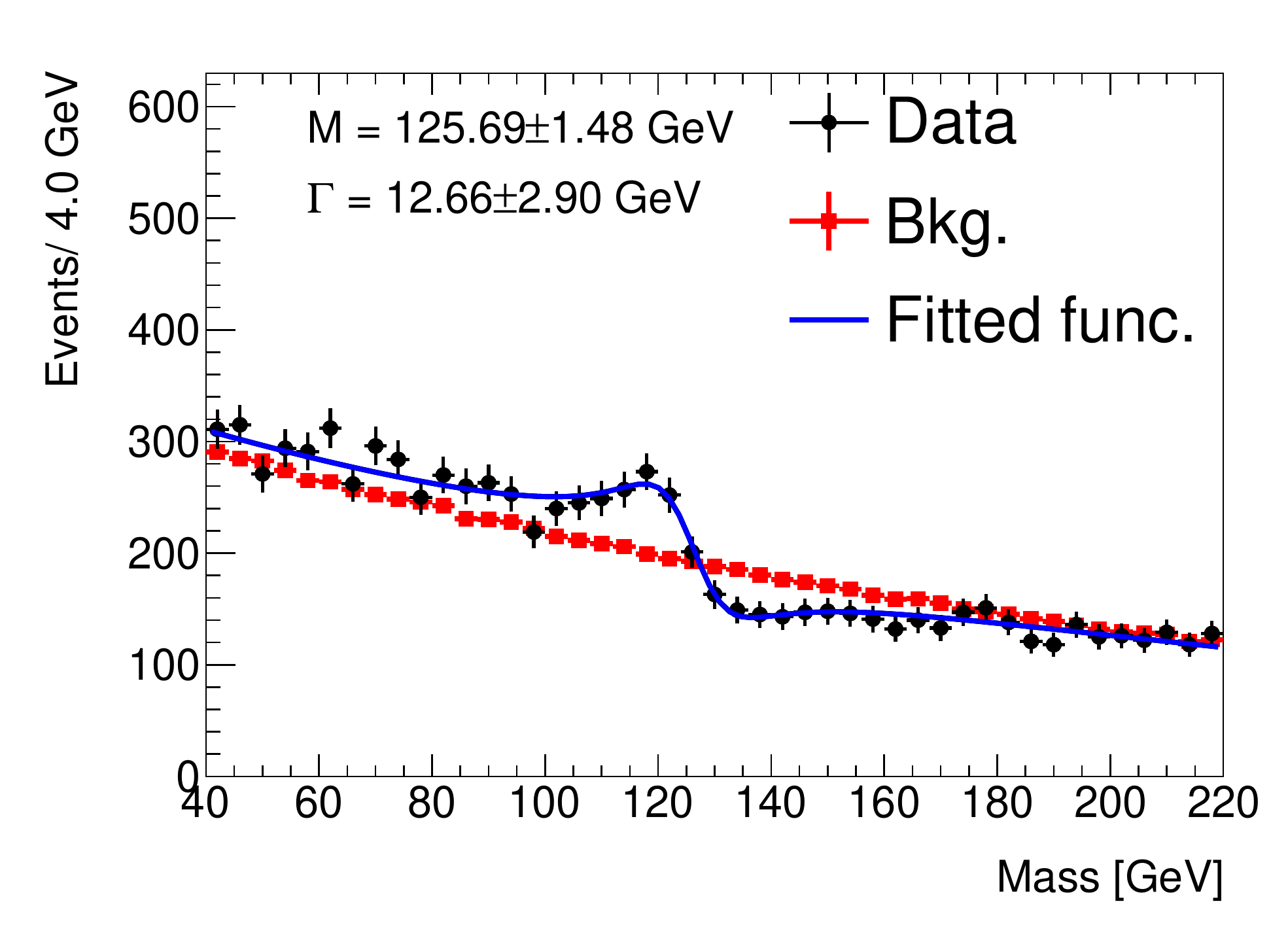}
    \includegraphics[width=0.32\textwidth]{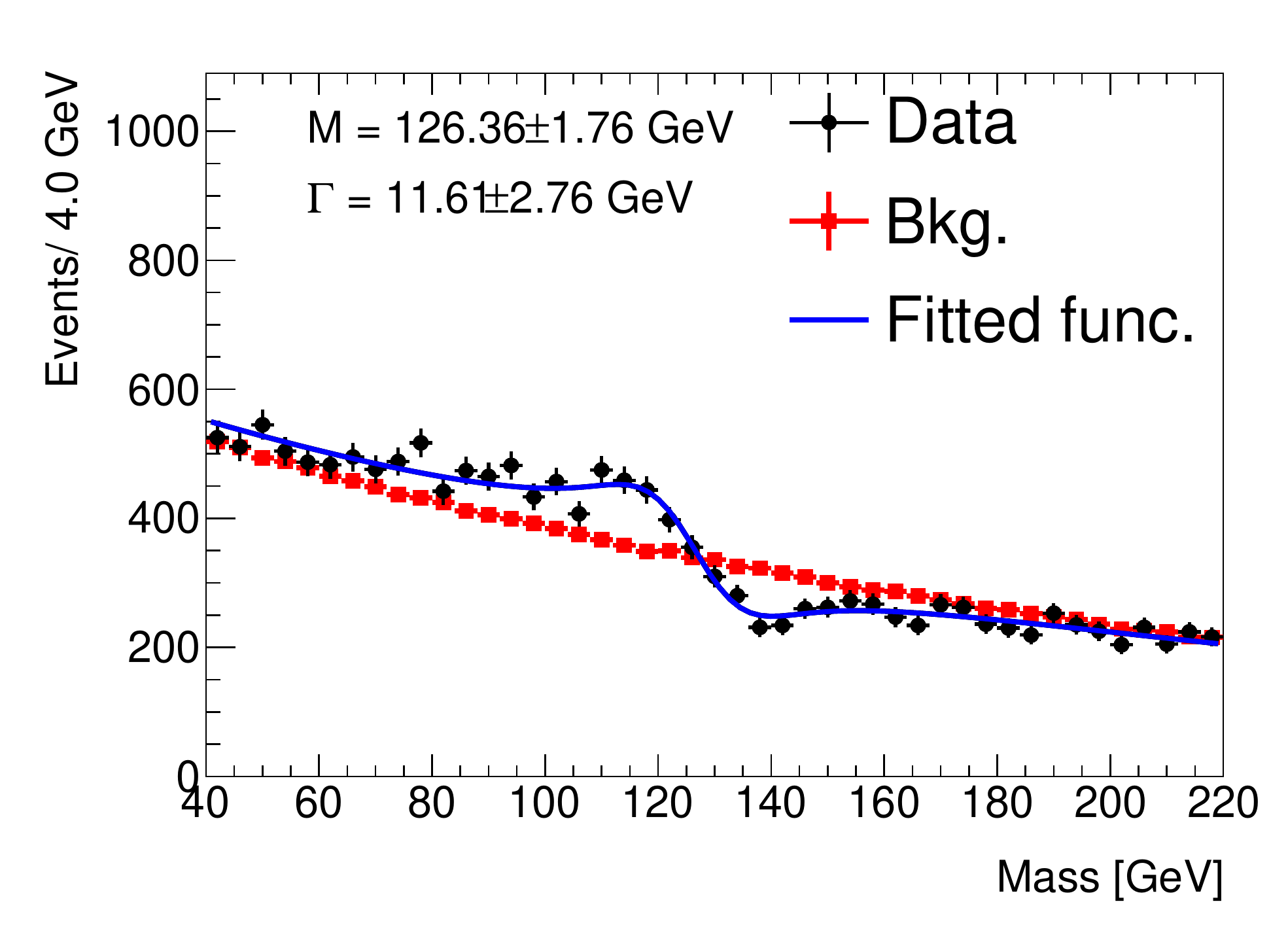}
    \includegraphics[width=0.32\textwidth]{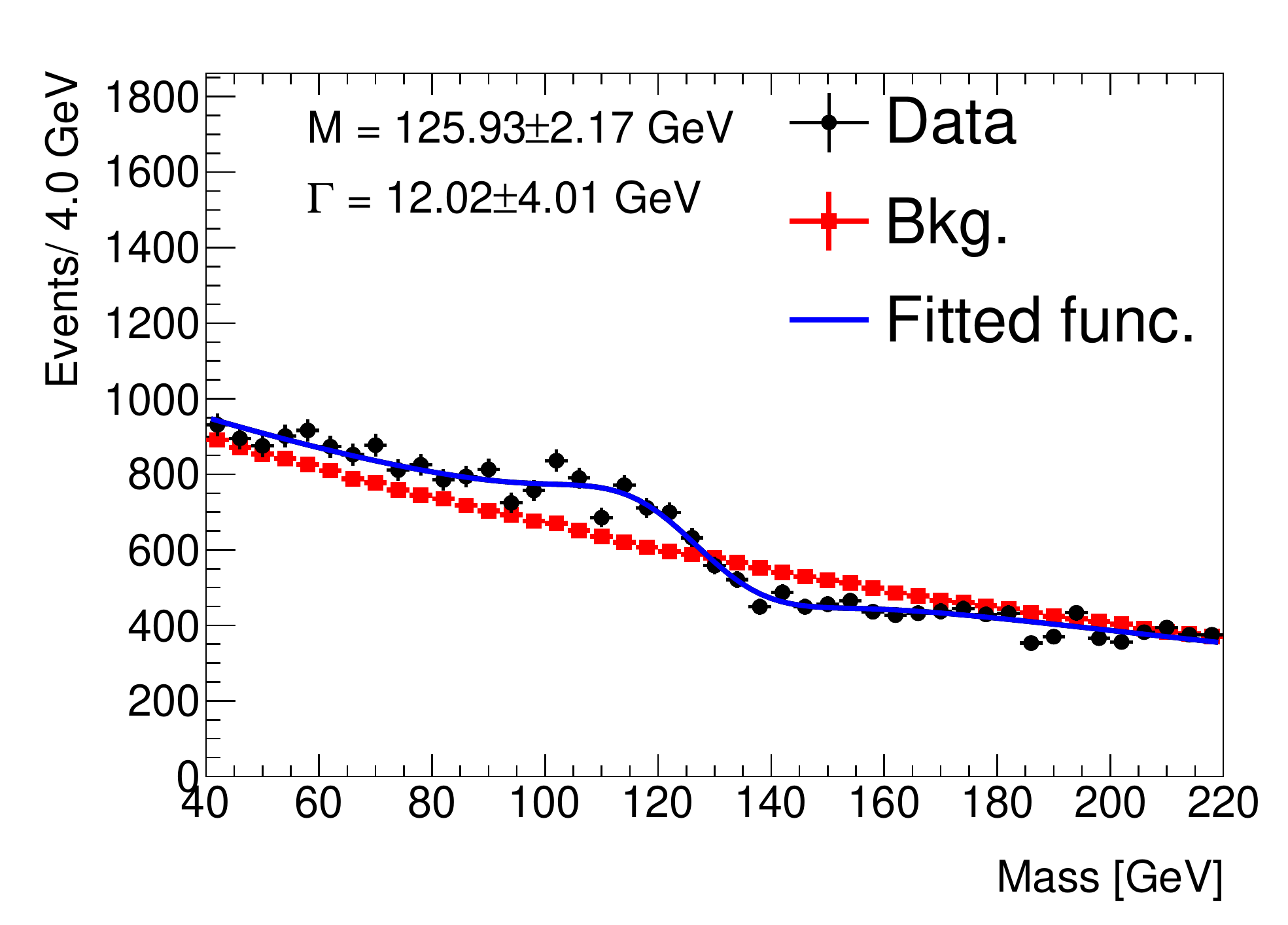}
    \caption{\label{fig:funcfit1}
        (color online) 
        Fitting results for three resolutions, 3~GeV (left), 6~GeV (middle) and 10~GeV (right). The black dots with error bar represent the data sample. The red squares represent the background sample. The blue curves represent the fit results. 
    }
\end{figure}

In this section, the mass distribution for the phase angle $\delta=180^\circ$ in last subsection is smeared out with three resolutions, 3~GeV, 6~GeV and 10~GeV, respectively. These distributions are shown in Fig.~\ref{fig:funcfit1} as well as the fitting results using the explicit signal model. 

\begin{figure}
    \includegraphics[width=0.32\textwidth]{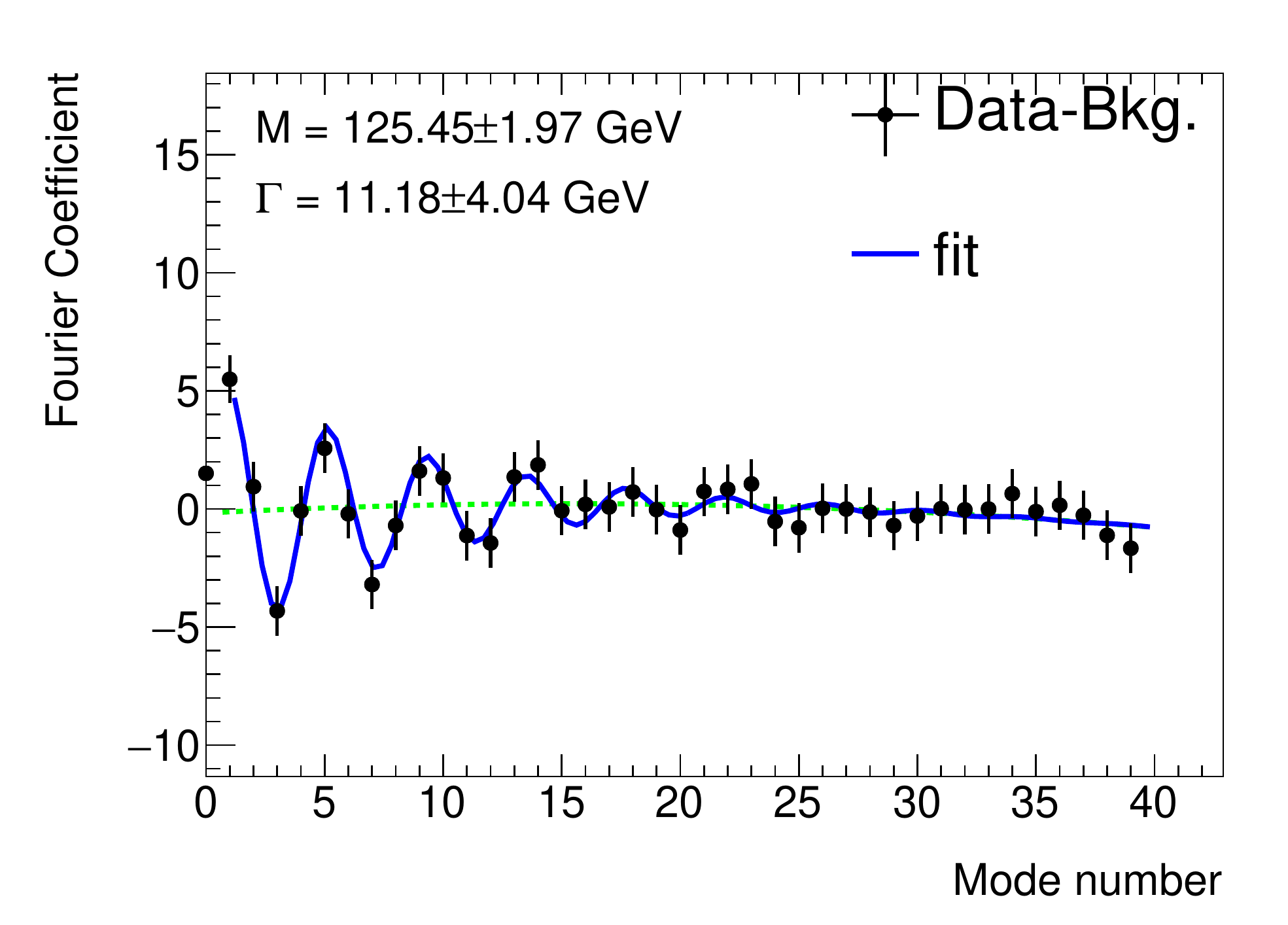}
    \includegraphics[width=0.32\textwidth]{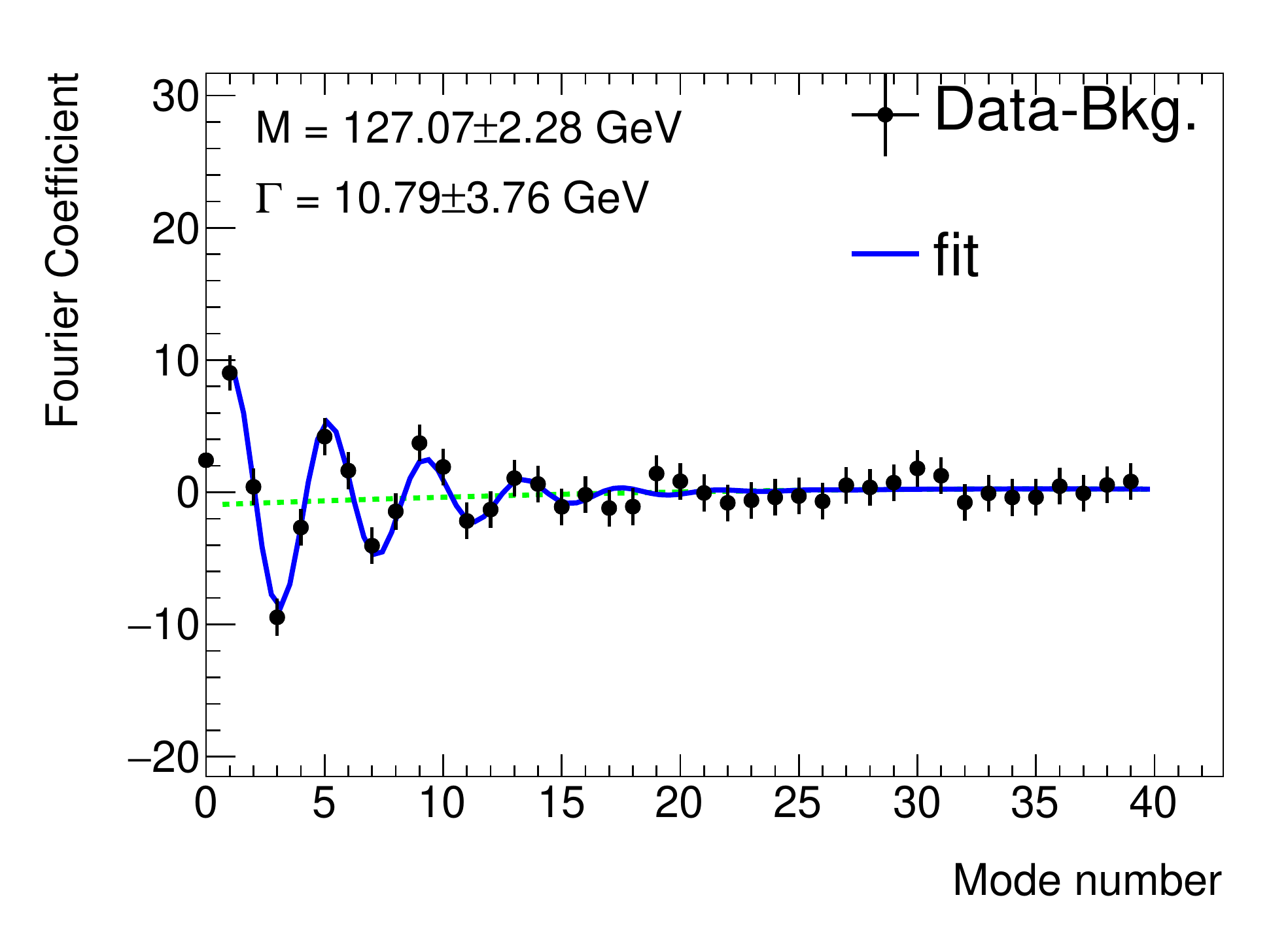}
    \includegraphics[width=0.32\textwidth]{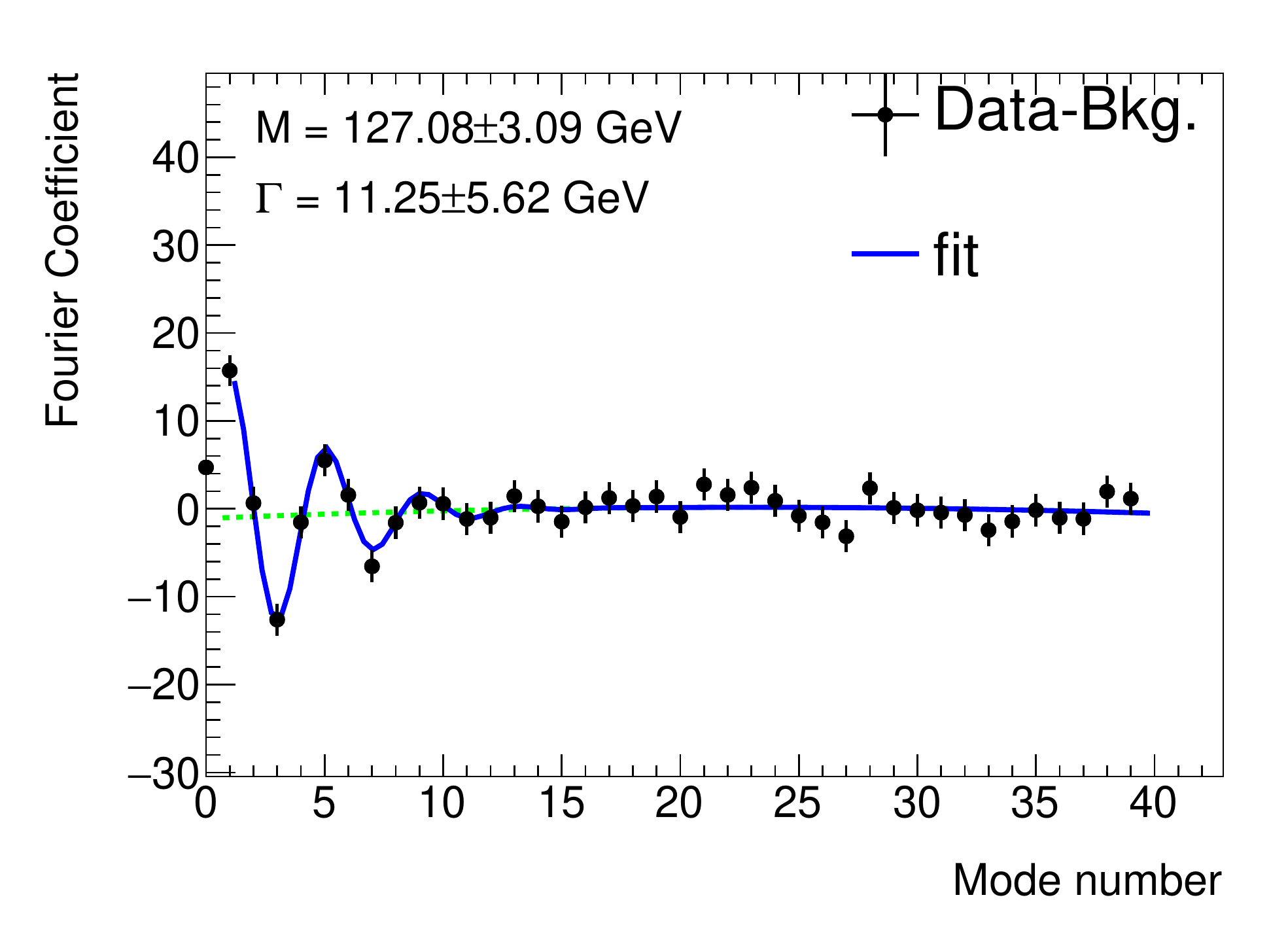}
    \caption{\label{fig:ckfitsigeff1}
        (color online) 
        Fitting results for three resolutions, 3~GeV (left), 6~GeV (middle) and 10~GeV (right). 
        The black dots with error bar represent the Fourier coefficients of the effective signal (data subtracting background). The blue curves represent the full fitting results. The green dashed curves represents the slow-varying contribution, namely, the polynominal part in Eq.~\ref{eq:fitfunc}. 
    }
\end{figure}

\begin{figure}
    \includegraphics[width=0.32\textwidth]{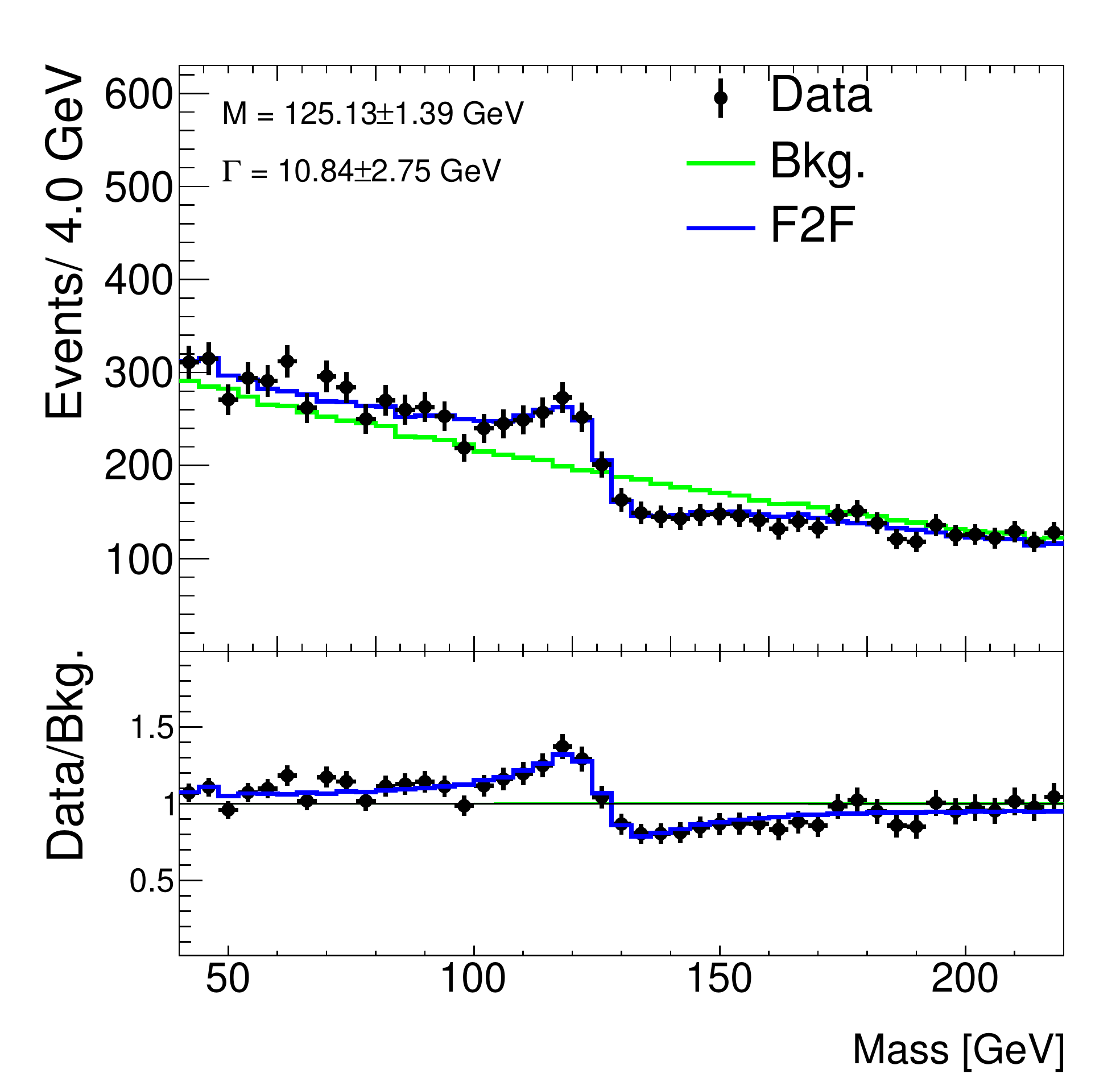}
    \includegraphics[width=0.32\textwidth]{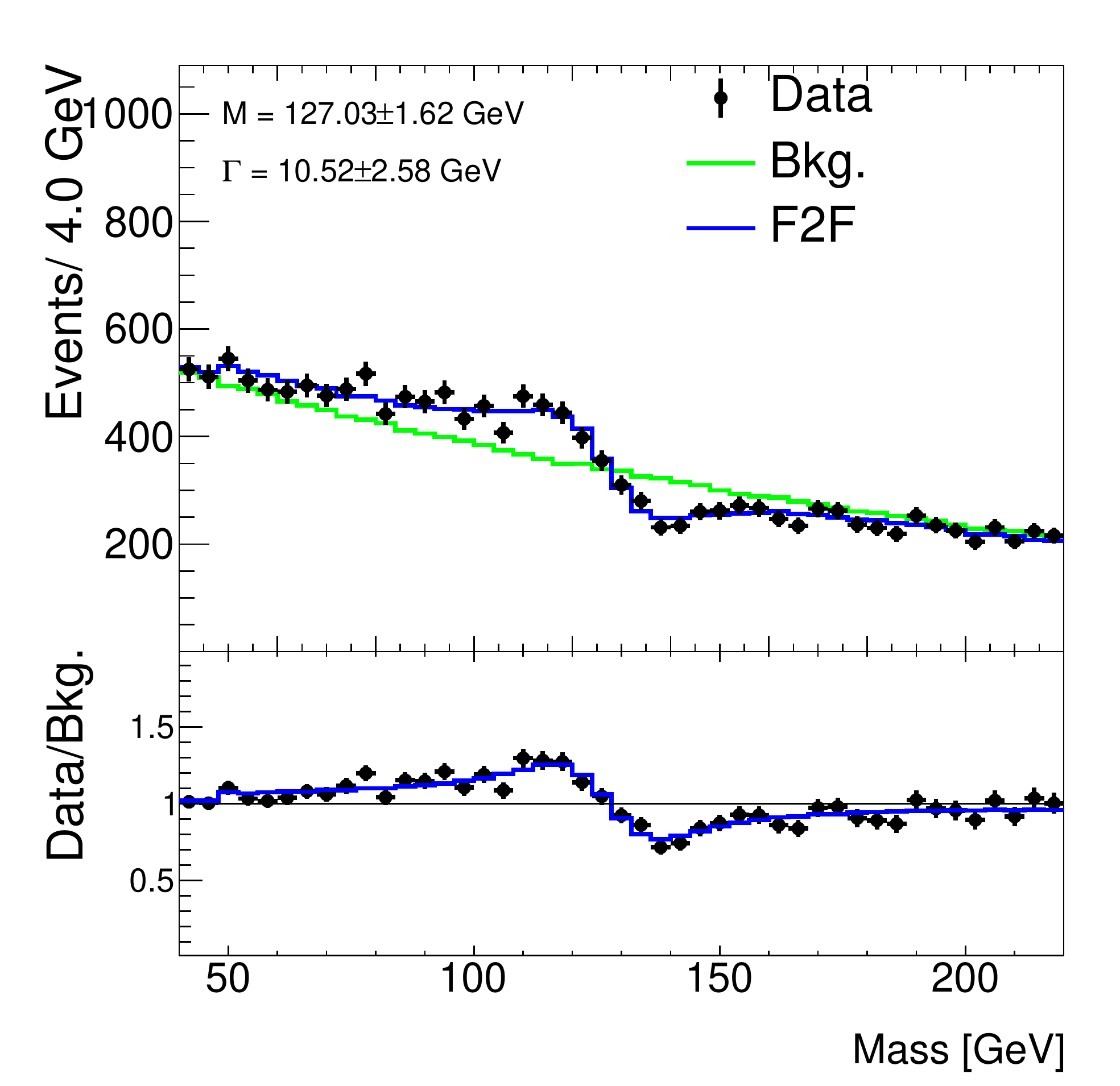}
    \includegraphics[width=0.32\textwidth]{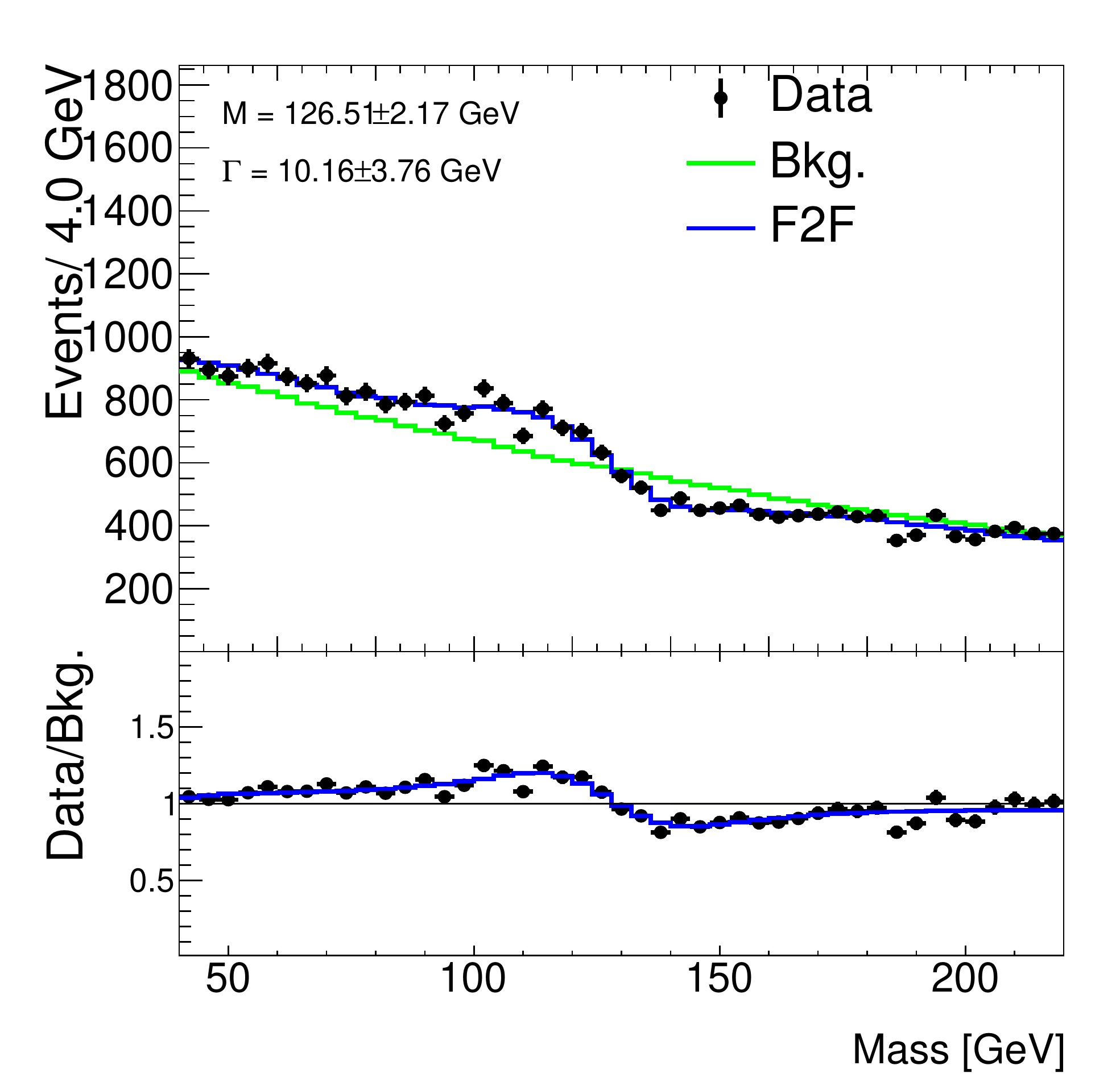}
    \caption{\label{fig:ckfitsigeff1p}
        (color online) 
        Fitting results for three resolutions, 3~GeV (left), 6~GeV (middle) and 10~GeV (right). 
        The black dots with error bar represent the Fourier coefficients of the effective signal (data subtracting background). The green histograms represent the background. The blue curves represent the full fitting results. 
    }
\end{figure}

The fitting results using the F2F method for setup1 and setup2 are shown in Fig.~\ref{fig:ckfitsigeff1} and Fig.~\ref{fig:ckfitsigeff1p}, respectively. 
Looking at Fig.~\ref{fig:ckfitsigeff1}, it is interesting to find that we see less oscillations as the detector resolution is bigger. This plot helps us to determine how many terms should be kept in the Fourier series.
The numerical results are summarized in Table~\ref{tab:fit1}. The estimated mass and width agree well with the inputs with similar precisions.

\begin{table}
    \caption{\label{tab:fit1}
        Summary of the fitting results on the mass and width in the case of imperfect detector resolution (unit: GeV). 
    }
    \begin{ruledtabular}
        \begin{tabular}{l | l l | l l | l l}
            & \multicolumn{2}{c}{Explicit signal model}  & \multicolumn{2}{c}{ F2F (setup1) } & \multicolumn{2}{c}{ F2F (setup2) }  \\
            \hline
            Resolution& Mass  & Width  & Mass  & Width  & Mass  & Width \\
            \hline
            3 & $125.69\pm1.48$ & $12.66\pm2.90$  & $125.45\pm1.97$ & $11.18\pm4.04$ & $125.13\pm1.39$ & $10.84\pm2.75$ \\
            6 & $126.36\pm1.76$ & $11.61\pm2.76$ & $127.07\pm2.28$ & $10.79\pm3.76$ & $127.03\pm1.62$ & $10.52\pm2.58$ \\
            10& $125.93\pm2.17$ & $12.02\pm4.01$ & $127.08\pm3.09$ & $11.25\pm5.62$ & $126.51\pm2.17$ & $10.16\pm3.76$ \\
        \end{tabular}
    \end{ruledtabular}
\end{table}

\begin{figure}
    \includegraphics[width=0.5\textwidth]{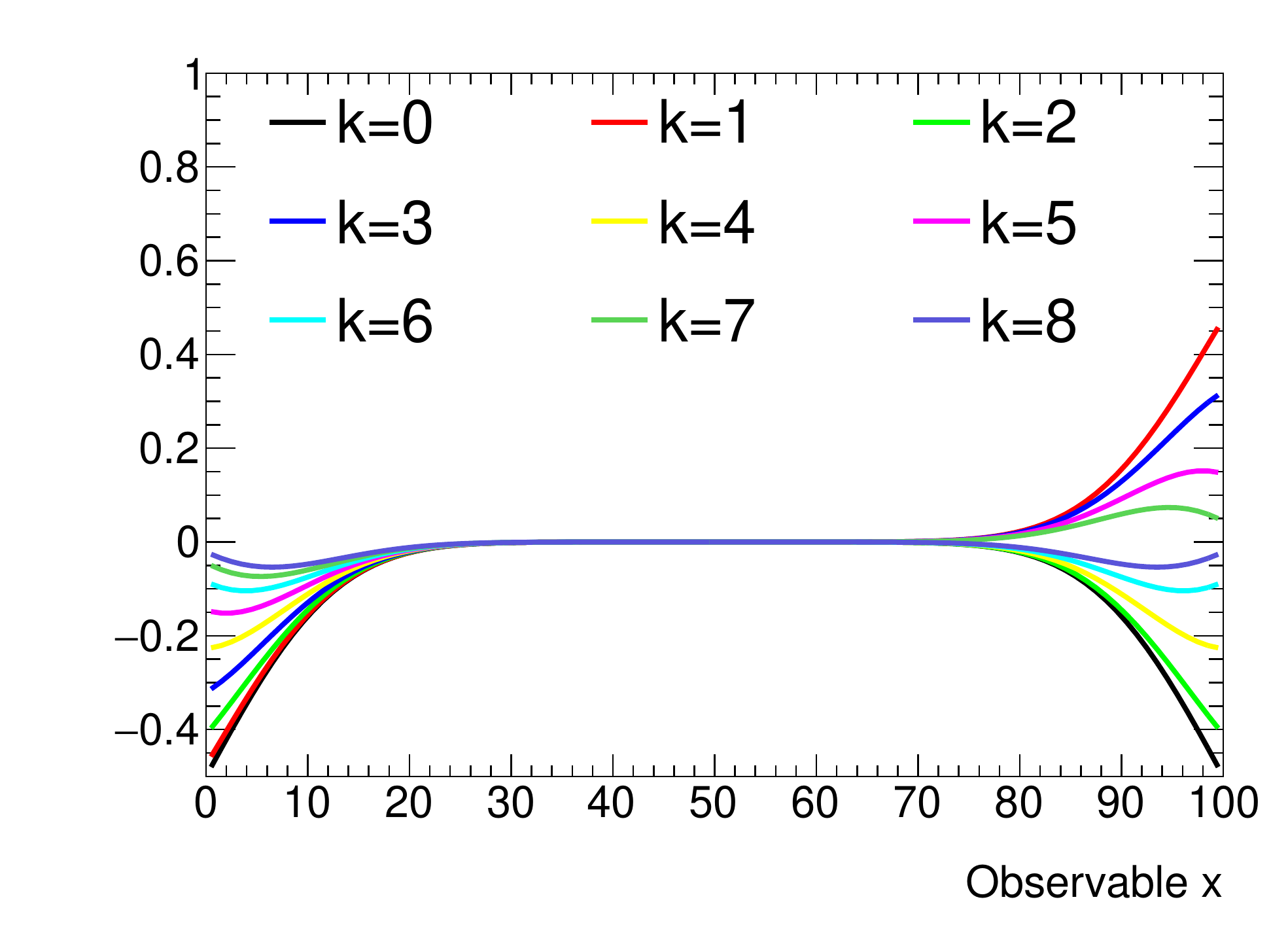}
    \caption{\label{fig:intcosgaus}
        (color online)
        The difference between the smeared distribution function with the convolution range $(-\infty,+\infty)$ and that with the convolution range $(0, 100)$ for different frequency modes. The resolution is 10.  
    }
\end{figure}

It is worth mentioning a subtle thing about smearing. By approximating a distribution function about an observable, $x$, by a cosine Fourier series in the range $(0,L)$, we actually assume the distribution function is symmetrical about $x=0$ and has a period $2L$. This makes the convolution with a Gaussian function very easy to calculate as shown in Eq.~\ref{eq:cksmear}, but also reveals a disadvantage of the Fourier series approximation. 
This is because the distribution in reality is probably neither symmetric about $x=0$, nor periodic. To check its potential effect, we restrict the convolution range from $(-\infty, +\infty)$ to $(0, L)$ in calculating the smeared out coefficients in Eq.~\ref{eq:cksmear}. Supposing $L=100$ and the resolution is 10 (the unit is non-relevant for the discussion), the smeared distribution differences between the two convolution ranges are shown in Fig.~\ref{fig:intcosgaus} for different frequency modes. As expected, the difference is only
significant around the edges. Therefore we should choose a range where the structure to be studied is far from both edges by a distance of at least 2 times the resolution (as observed in Fig.~\ref{fig:intcosgaus}). 

\section{Application~1: SM Higgs width measurement}\label{sec:higgswidth}
In this section, we apply the F2F method to the measurement of the width of the SM Higgs boson, $h$, using the 4 muon final state. We have two signal processes, $gg\to h \to ZZ^* \to 2\mu^+2\mu^-$ and $qq\rightarrow h \to ZZ^* \to 2\mu^+\mu^-$, where the former dominates. They are simulated using the MC generator gg2VV~\cite{gg2VV0,gg2VV1} and Madgraph5\_aMC@NLO~\cite{madgraph}, respectively. Correspondingly, we also have two coherent background processes, $gg\to 2\mu^+2\mu^-$ and $qq\to 2\mu^+2\mu^-$, where the Higgs boson is not involved and the latter one dominates.

\begin{figure}
    \includegraphics[width=0.32\textwidth]{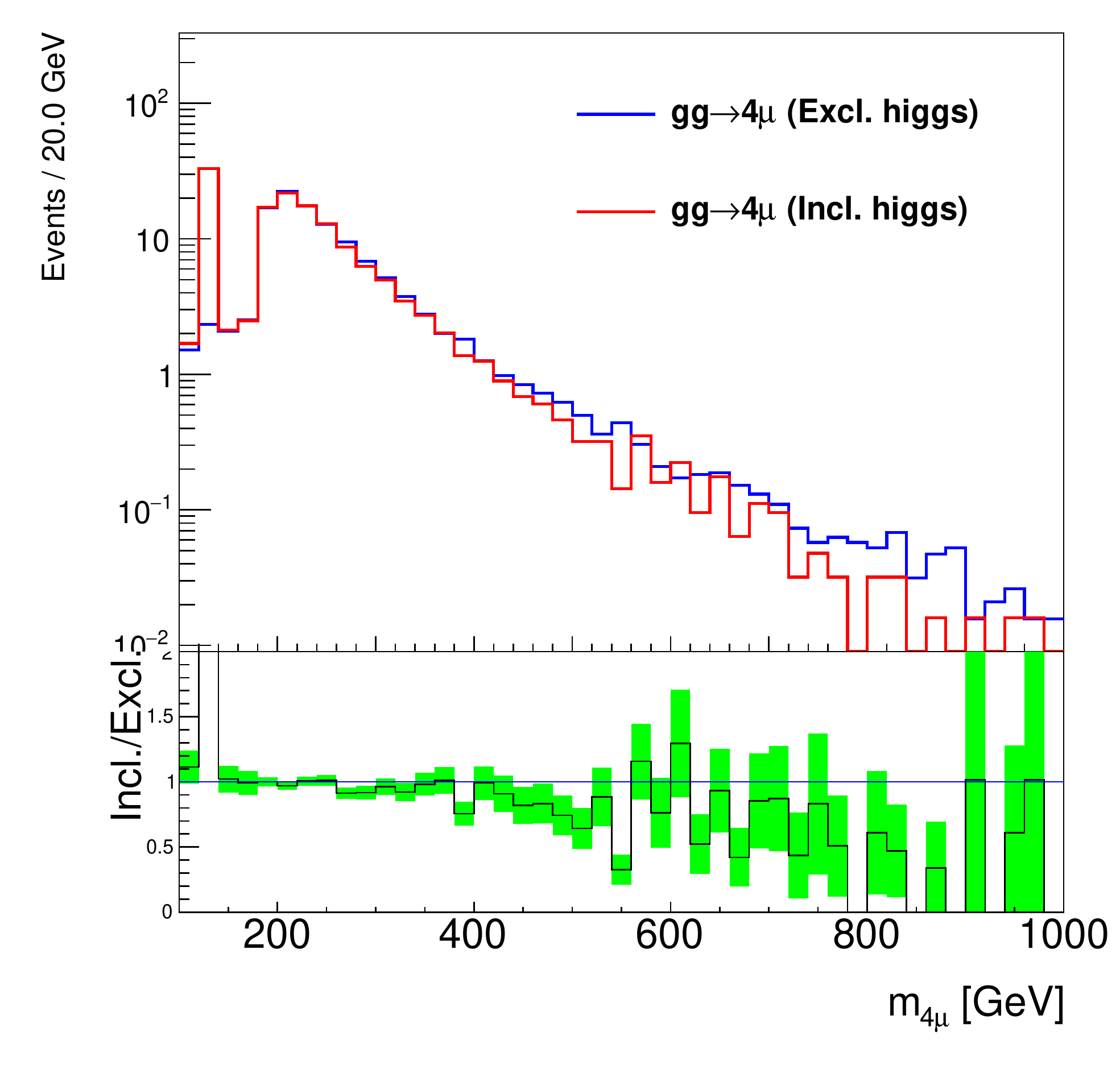}
    \includegraphics[width=0.32\textwidth]{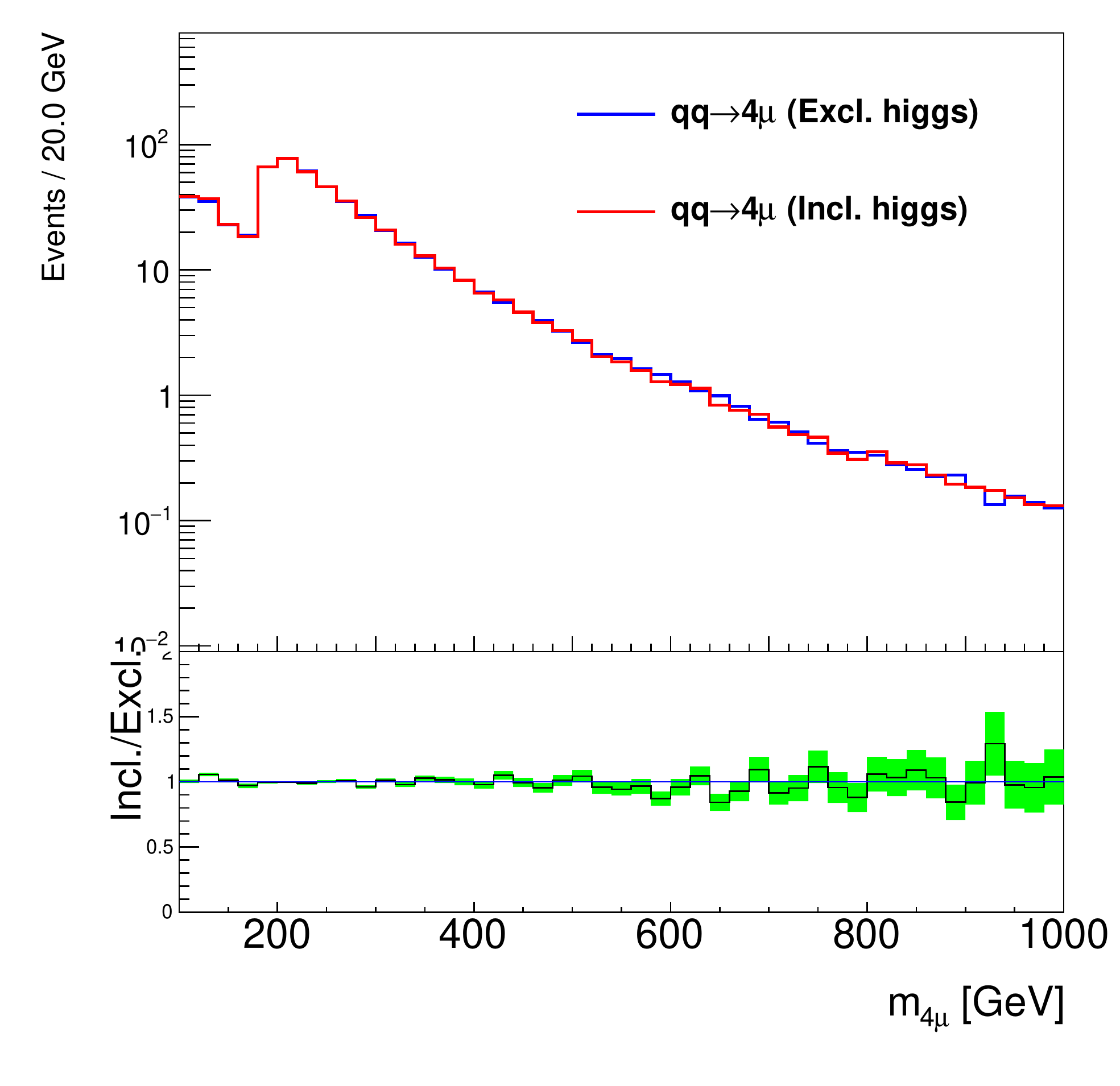}
    \includegraphics[width=0.32\textwidth]{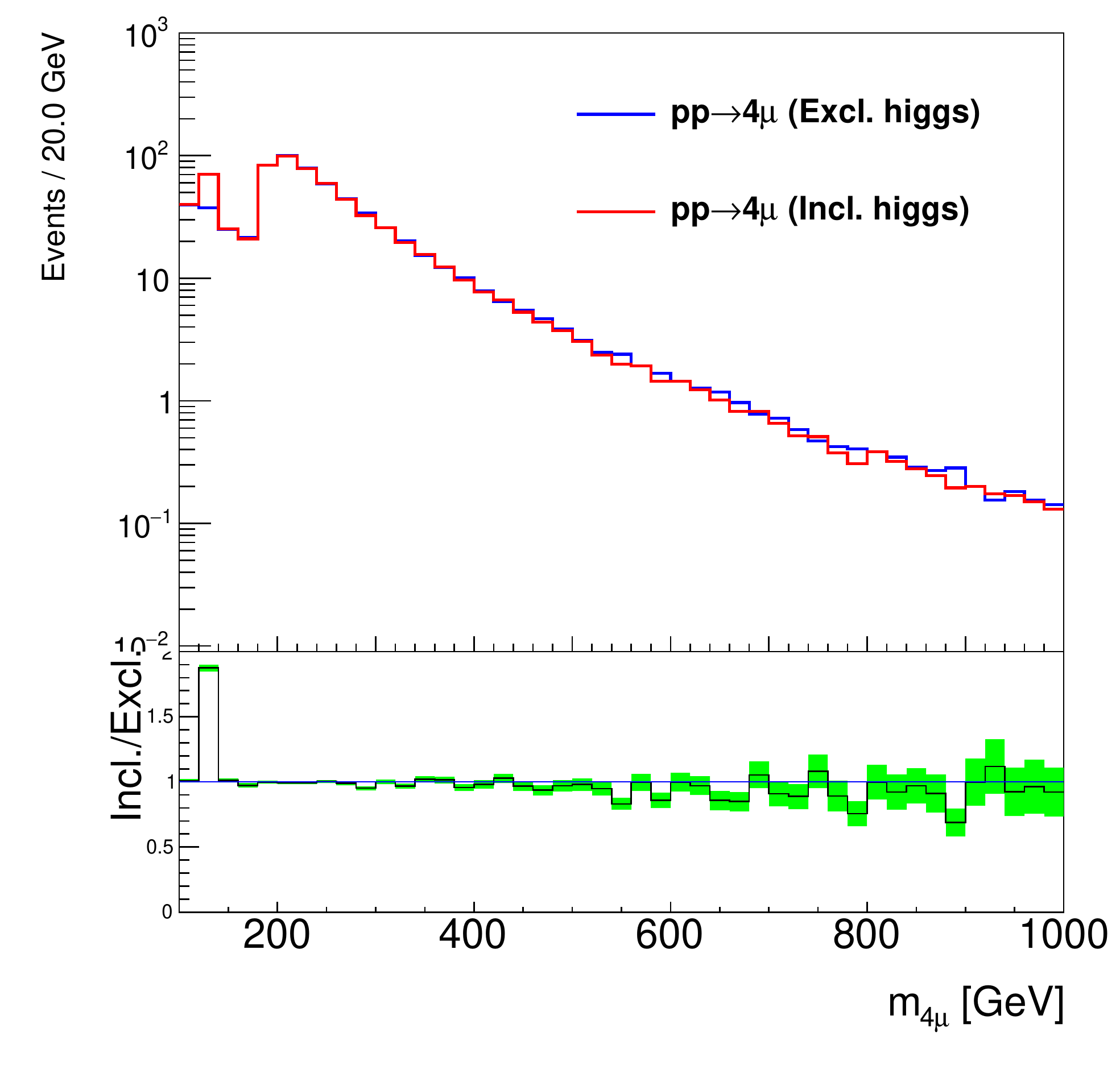}
    \caption{\label{fig:higgswidth_m4l}
        (color online)
        The four-muon mass distribution from the process $gg\to4\mu$ (L), $qq\to4\mu$ (M) and their combination $pp\to4\mu$ (R). The red (blue) histogram represents the distribution with (without) including the Higgs boson in the upper pad while their ratio is shown in the lower pad. The green band in the lower pad indicates the background MC uncertainty (the uncertainty itself is not considered in the fit).
    }
\end{figure}

The events are selected with similar but simpler conditions as in Ref.~\cite{ATLAS_higgswidth, CMS_higgswidth}. Muons are required to have transverse momentum, $\pT$, larger than 3~GeV and absolute pseudo rapidity, $|\eta|$, less than $2.6$ in the event generation. We further require there are at least 1 muons with $\pT>20$~GeV and at least 2 muons with $\pT>10$~GeV. The invariance mass of any $\mu^+\mu^-$ pair, $m_{\mu^+\mu^-}$, is required to be greater than 10~GeV. The pair of $\mu^+\mu^-$ with
closest mass to the $Z$ boson (denoted by $\mu^+\mu^-(Z)$) is required to satisfy $12<m_{\mu^+\mu^-(Z)}<120$~GeV while the other pair (denoted by $\mu^+\mu^-(Z^*)$) to satisfy $m_{\mu^+\mu^-(Z^*)}>40$~GeV if $m_{\mu^+\mu^-(Z)}<40$~GeV. 
For simplicity, we do not consider other non-coherent backgrounds and do not perform detector simulation. But a Gaussian smearing with a resolution 1.19~GeV~\cite{CMS_4muresolution} is applied to the invariant mass of the 4-muon final state, $m_{4\mu}$. The mass distributions for different processes with or without the Higgs resonance are shown in Fig.~\ref{fig:higgswidth_m4l} normalized to a luminosity of 80~fb$^{-1}$. We can see that the interference effect is destructive at high-mass region.

\begin{figure}
    \includegraphics[width=0.32\textwidth]{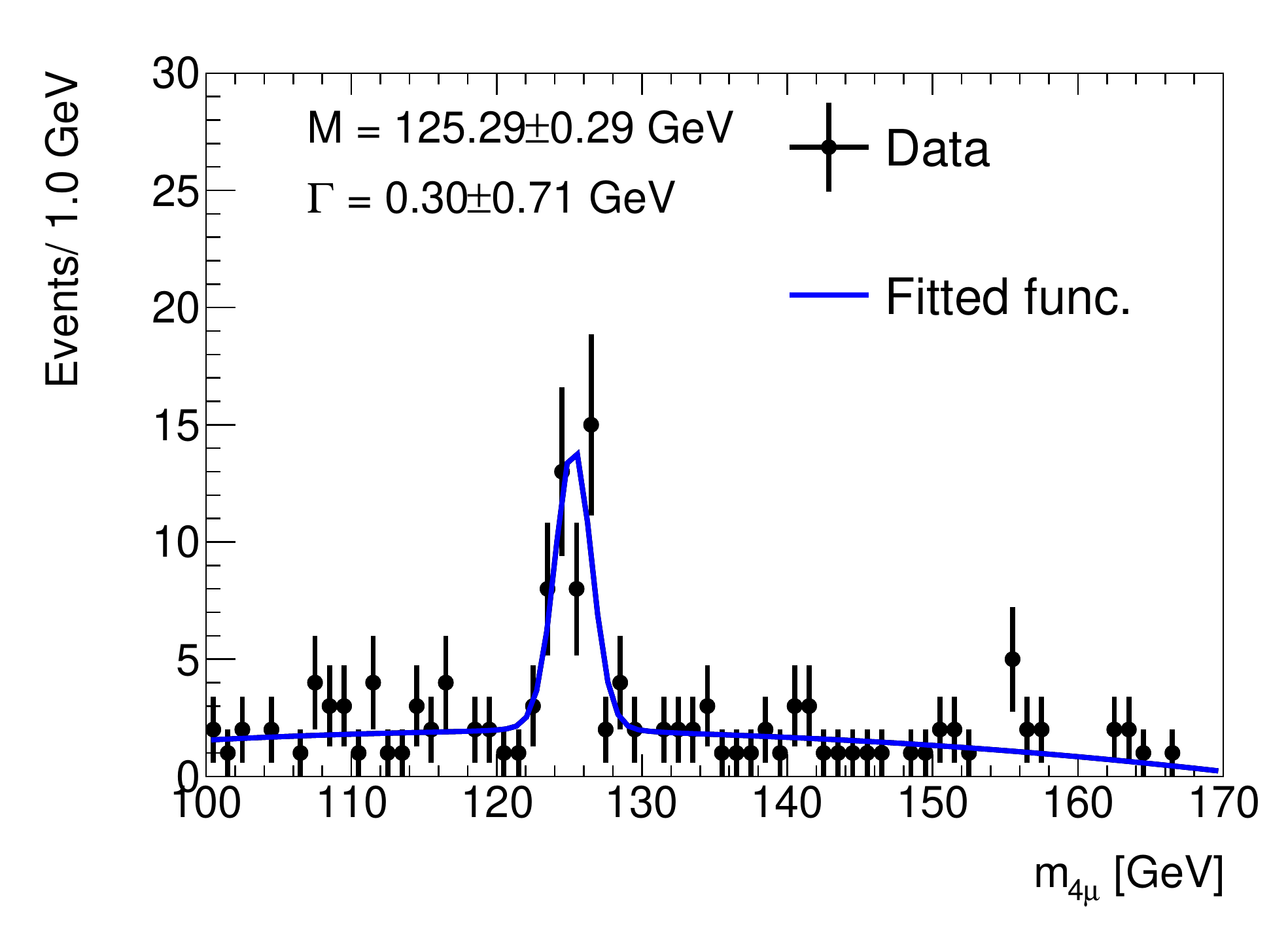}
    \includegraphics[width=0.32\textwidth]{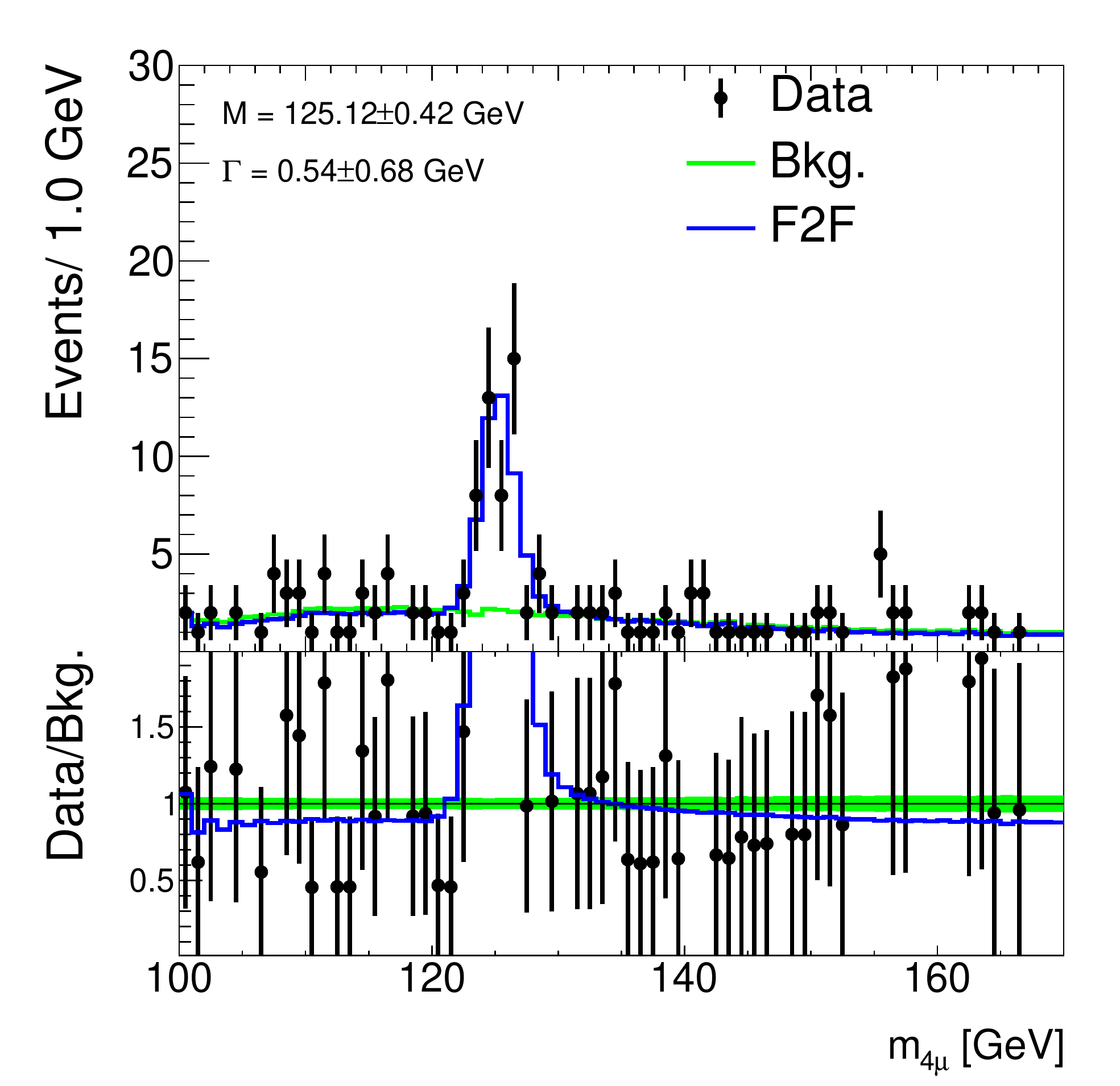}
    \includegraphics[width=0.32\textwidth]{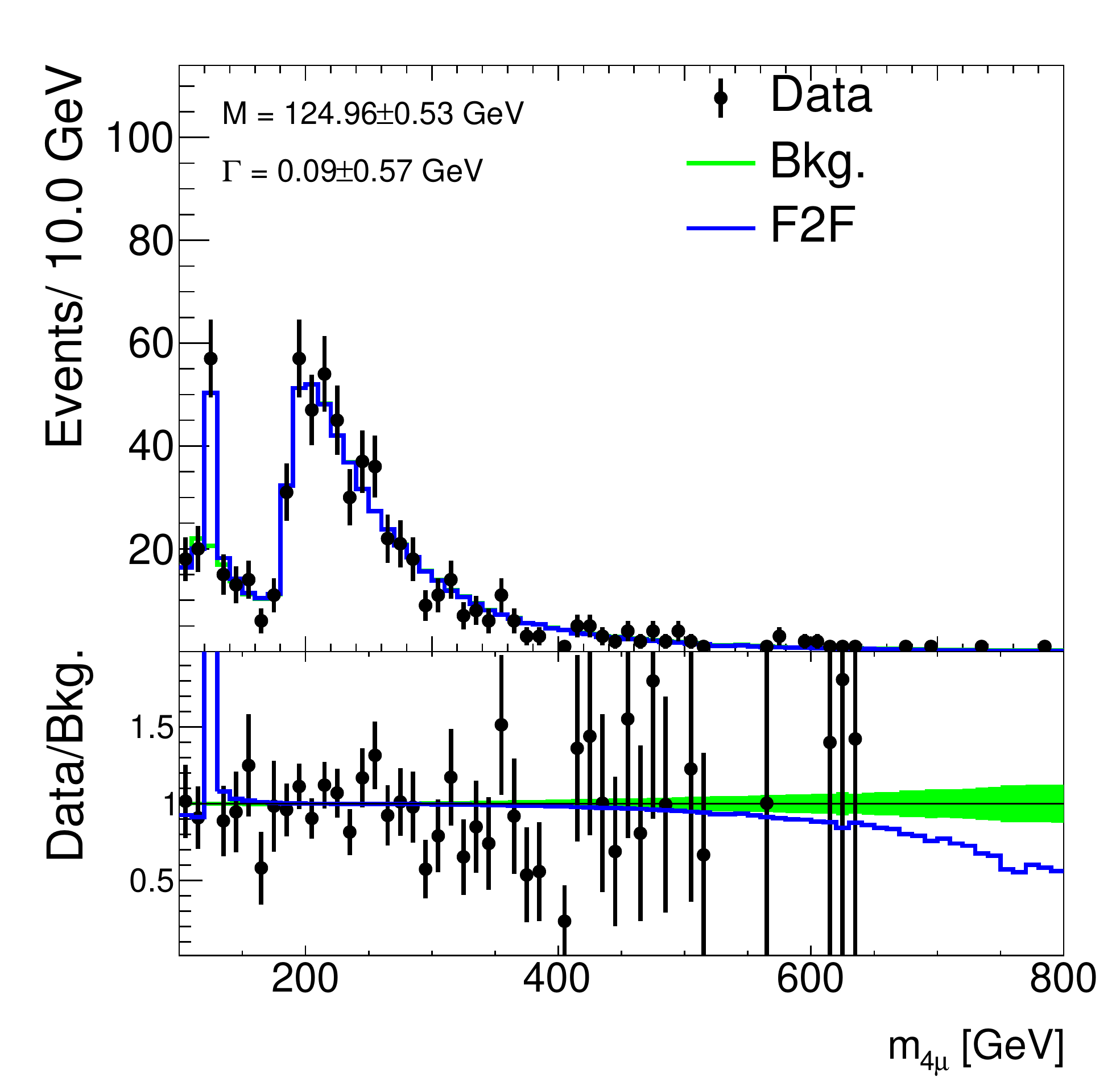}
    \caption{\label{fig:higgswidth_toy0}
        (color online) 
        Nominal fit in the on-shell region (L), F2F fit in the on-shell region (M) and F2F fit in the long region (R). The black dots with error bar represent data; the green histograms in the middle and right plots represent the background; the blue curves represent the best fit. In the middle and right plots, the green band in the lower pad indicates the background MC uncertainty (the uncertainty itself is not considered in the fit). 
    }
\end{figure}

To investigate the performance, the F2F method is applied in two regions, namely, the on-shell region ($100<m_{4\mu}<170$~GeV) and the long region ($100<m_{4\mu}<800$~GeV). In the on-shell region, a likelihood fit without considering the interference effect is also performed. We call it ``nominal fit'' throughout this section. 2000 toy data samples are generated to obtain the expected sensitivity on the width measurement. 

\begin{figure}
    \includegraphics[width=0.32\textwidth]{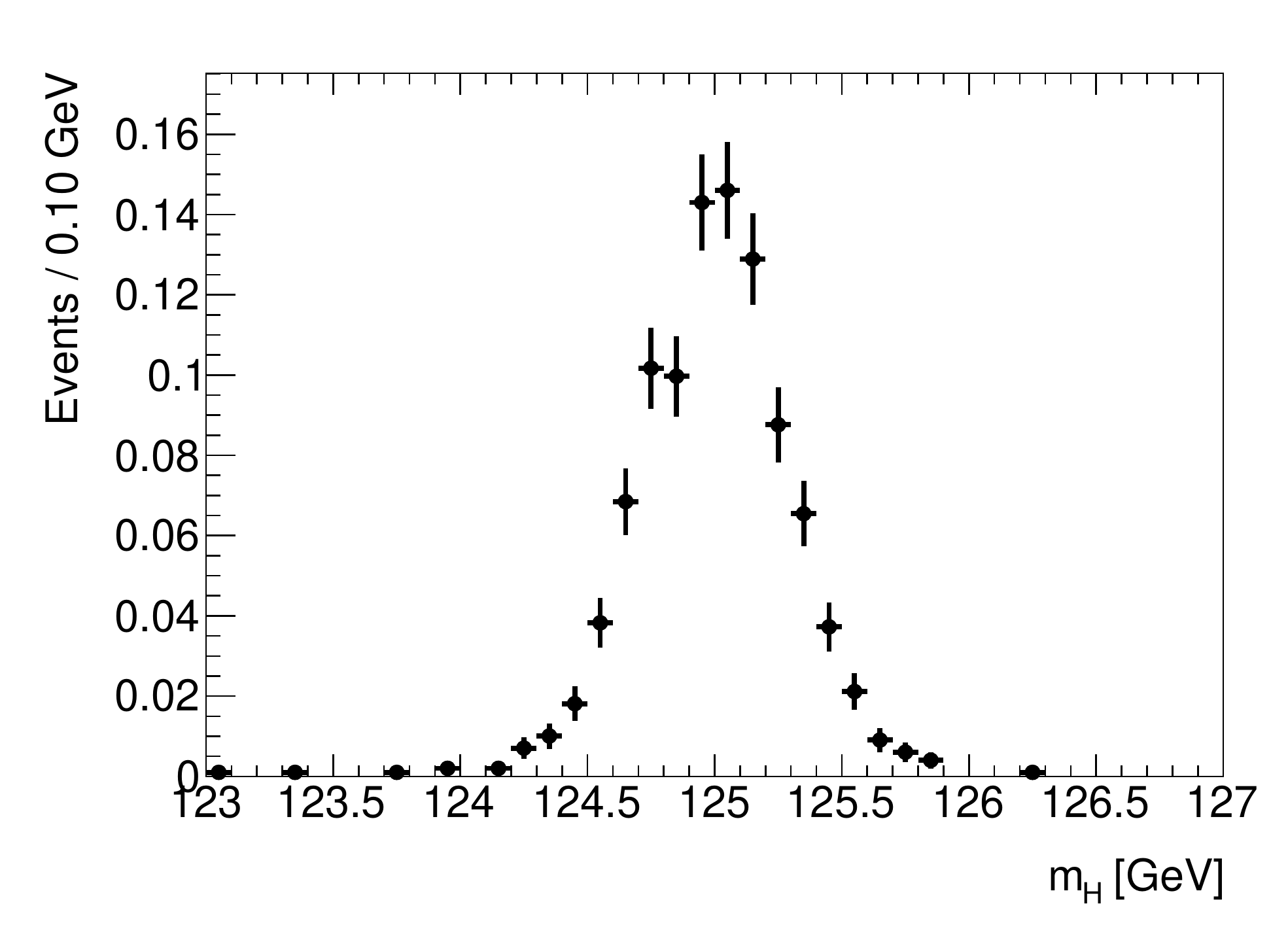}
    \includegraphics[width=0.32\textwidth]{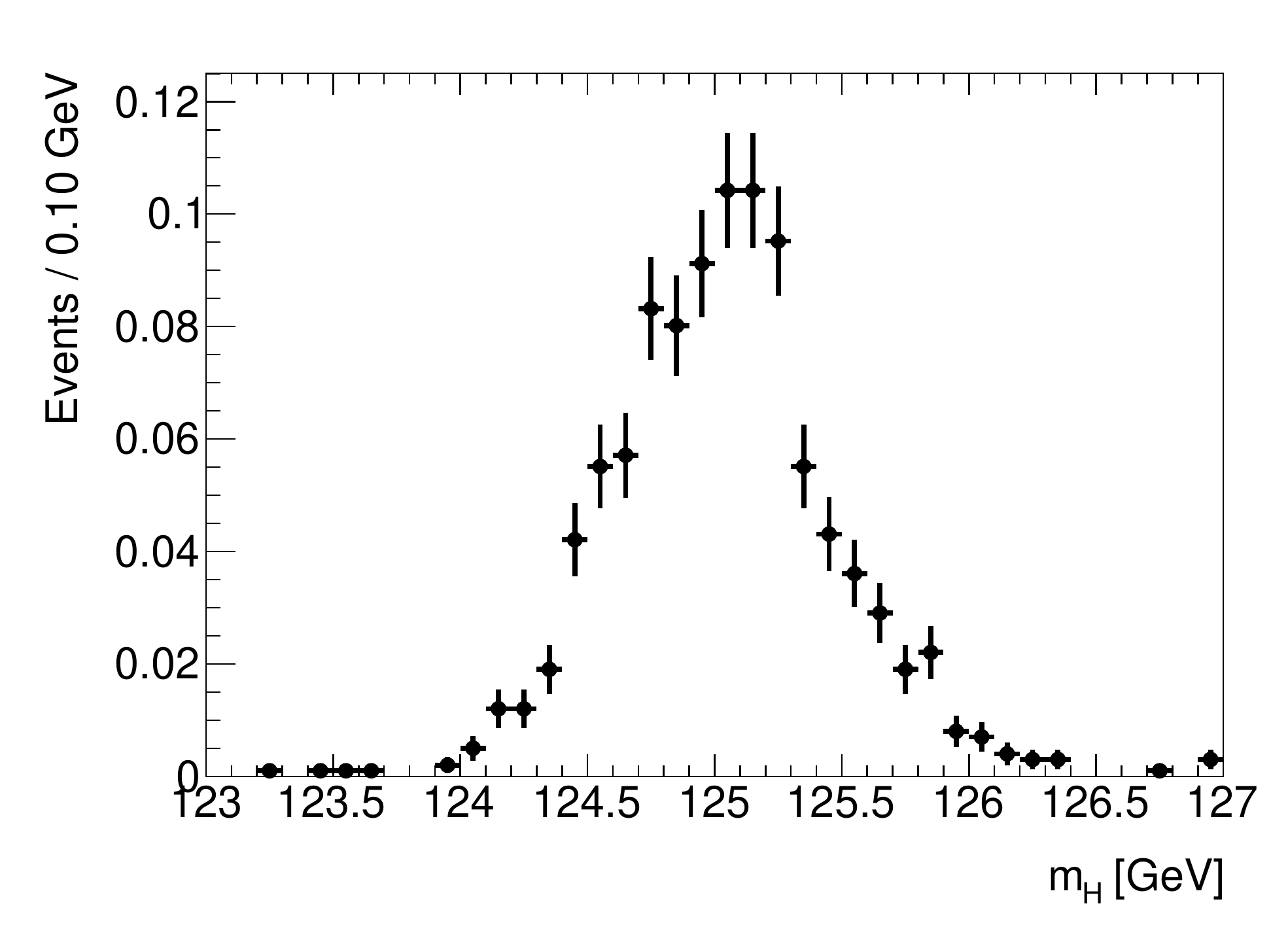}
    \includegraphics[width=0.32\textwidth]{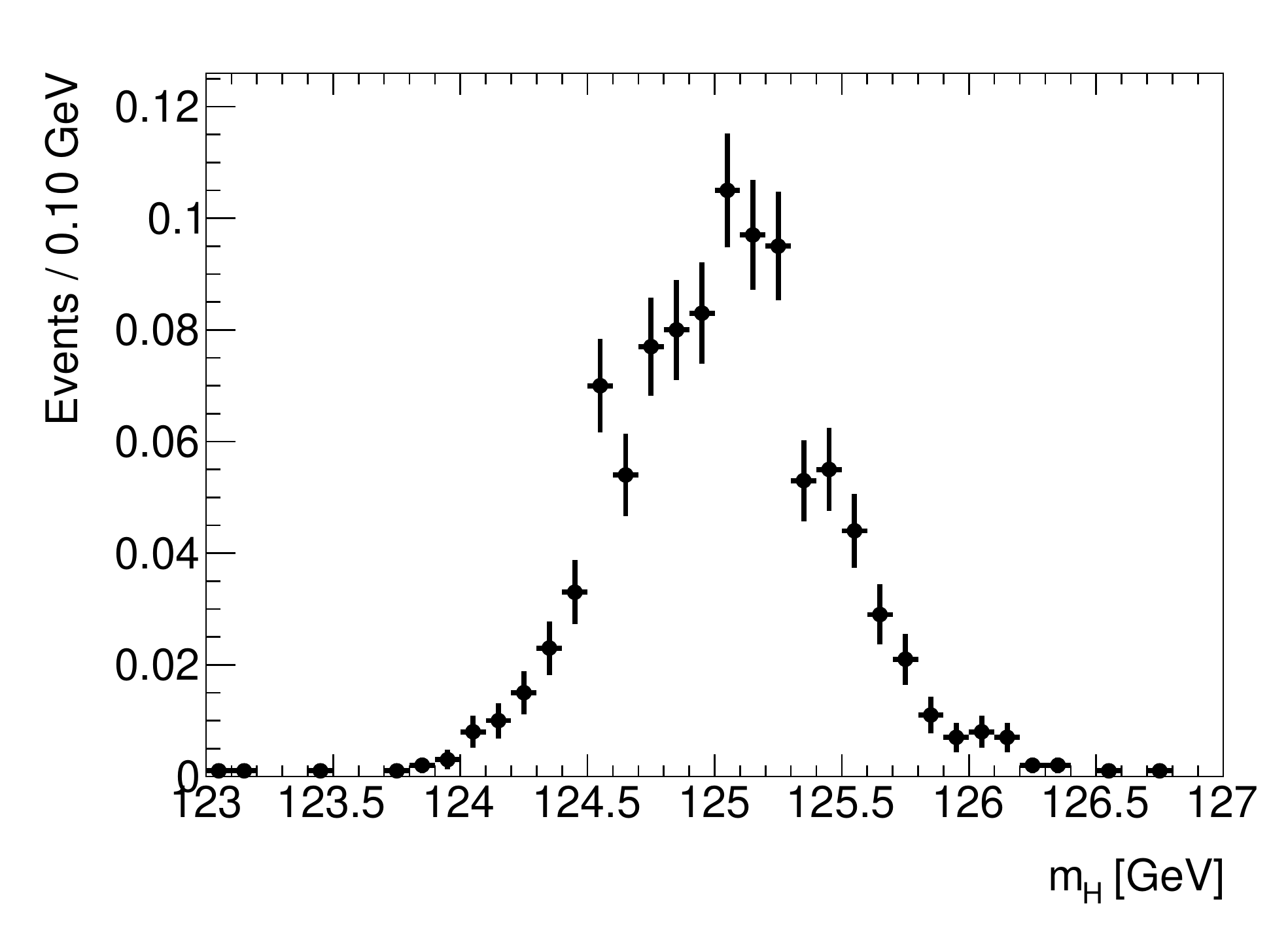}
    \caption{\label{fig:higgswidth_mass}
        (color online) 
        Distribution of the Higgs mass from the nominal fit in the on-shell region (L), F2F fit in the on-shell region (M) and F2F fit in the long region (R). All distributions are normalized to unit area.
    }
\end{figure}

\begin{figure}
    \includegraphics[width=0.32\textwidth]{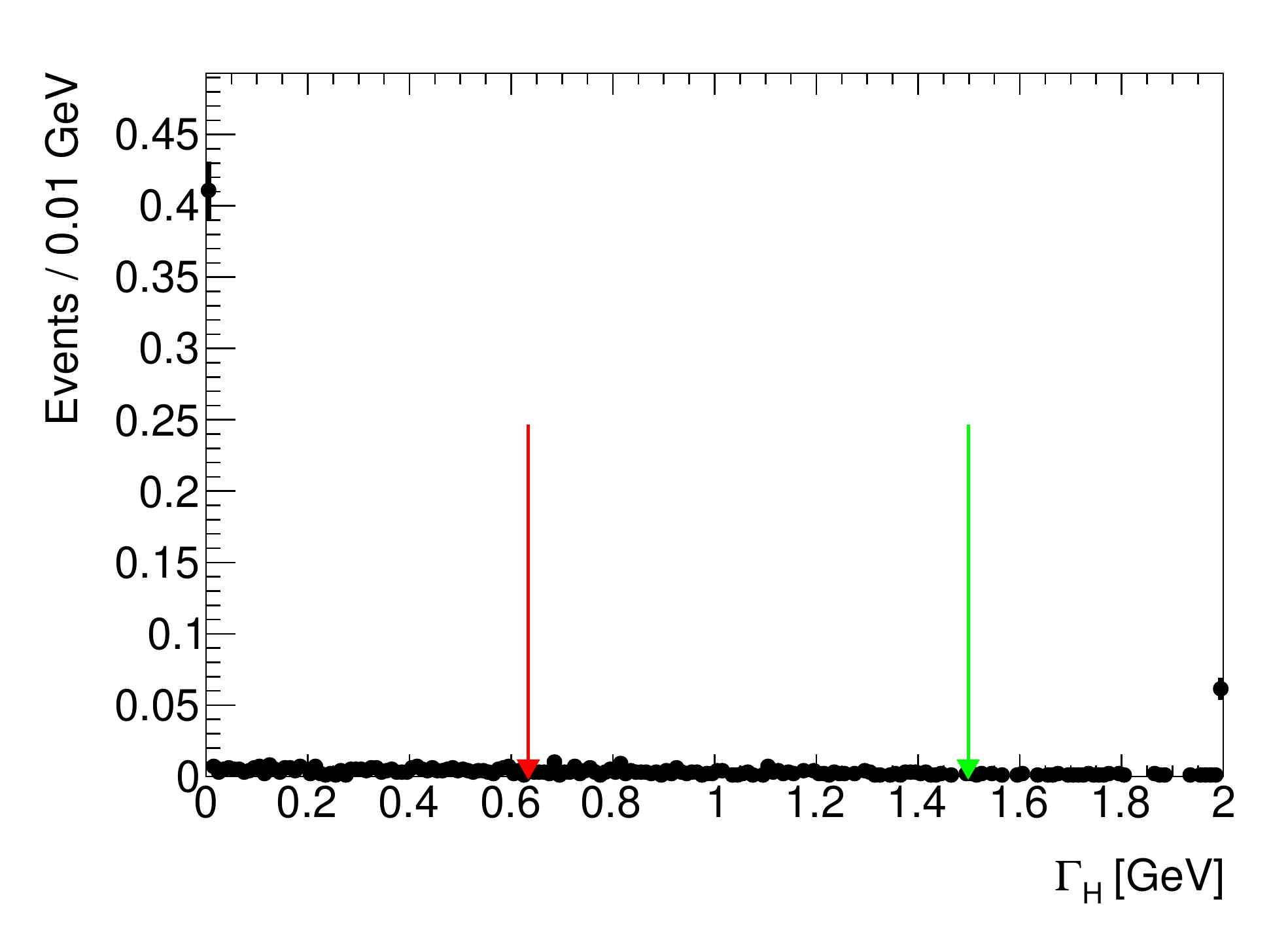}
    \includegraphics[width=0.32\textwidth]{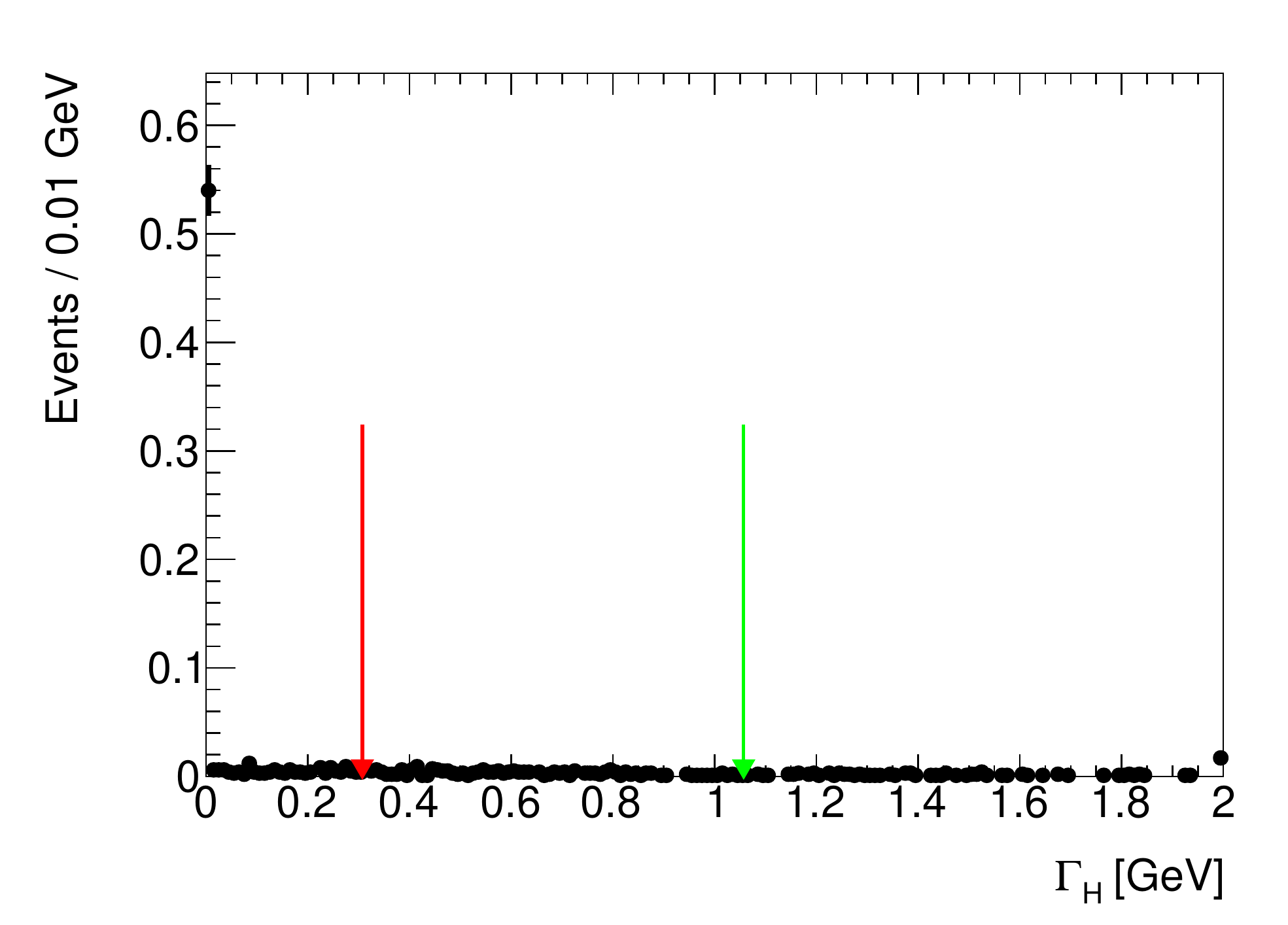}
    \includegraphics[width=0.32\textwidth]{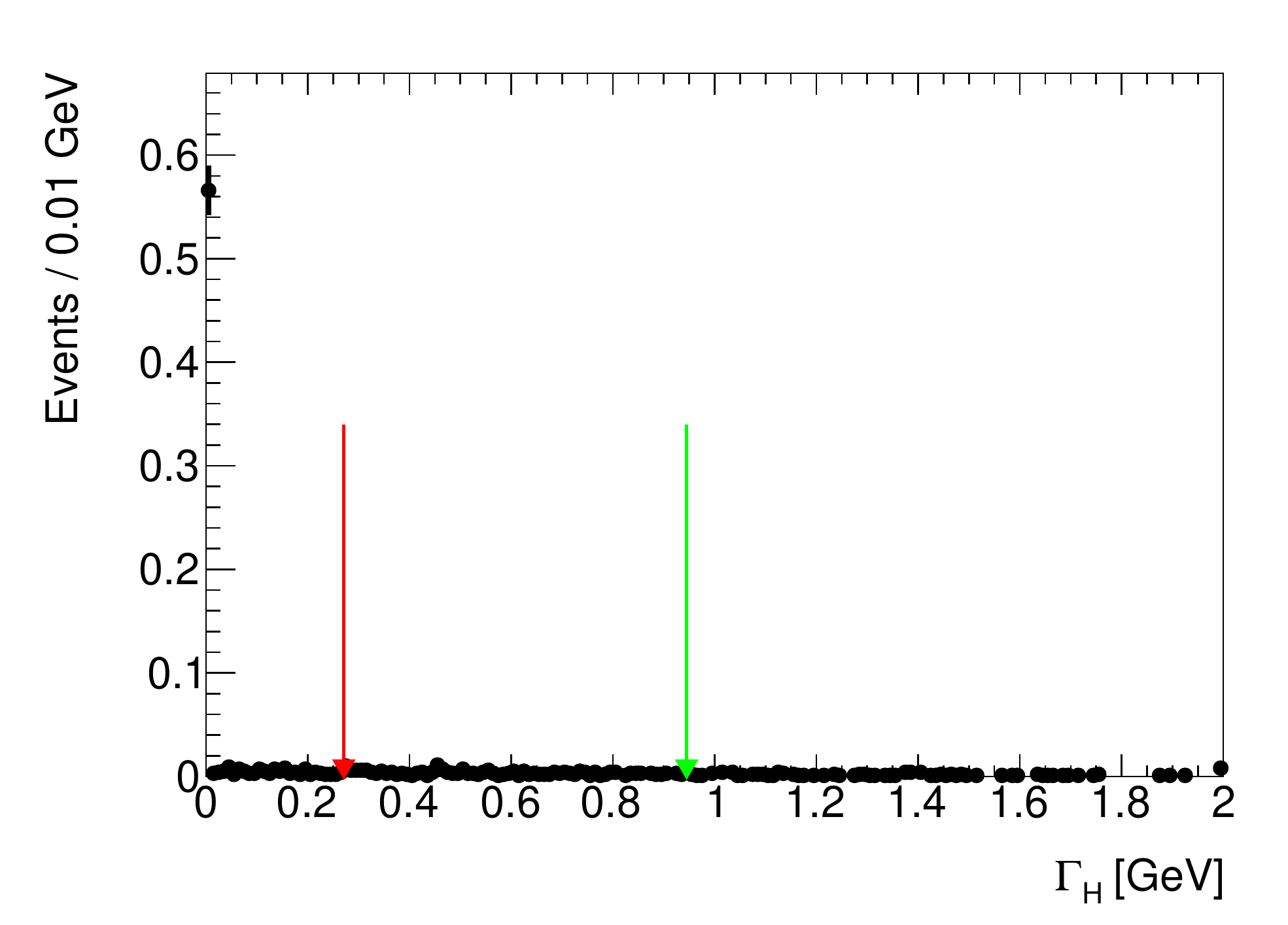}
    \caption{\label{fig:higgswidth_width}
        (color online) 
        Distribution of the Higgs width from the nominal fit in the on-shell region (L), F2F fit in the on-shell region (M) and F2F fit in the long region (R). All distributions are normalized to unit area. The red and green arrows represent the upper limits at 68.3~\% and 90~\% confidence level, respectively. 
    }
\end{figure}

Figure~\ref{fig:higgswidth_toy0} shows the three fitting results to a toy data sample. The distribution of the best-fit Higgs mass and width from 1000 toys is shown in Fig.~\ref{fig:higgswidth_mass} and Fig.~\ref{fig:higgswidth_width}, respectively. We can see that the distribution is Gaussian for the mass while non-Gaussian for the width. Especially, a big fraction of toys give very small higgs width, which is why the first bin is very high in Fig.~\ref{fig:higgswidth_width}. To
illustrate the sensitivity, the upper limits corresponding to 68.3~\% Confidence Level (C.L.) and 90~\% C.L. are found and shown by arrows in Fig.~\ref{fig:higgswidth_width}. 

All numerical results are summarized in Table~\ref{tab:higgswidth}. The nominal fit has a better sensitivity on the Higgs mass but its sensitivity on the width is much worse than the F2F fit (even just in the on-shell region). The F2F fit in the long region has the best sensitivity on the width measurement. It shows that the new method is promising for the SM Higgs width measurment. 

In the end of this section, it is worth noting that the long region only uses events with $m_{4\mu}$ up to 800~GeV. In real measurements~\cite{ATLAS_higgswidth,CMS_higgswidth}, events with the four-lepton mass up
to 2000~GeV or more are used. To use longer region, we have to prepare hugh MC samples so that the tiny interference effect is visible against the MC uncertainty itself. The traditional off-shell method is also not compared in this paper due to its complicity~\cite{ATLAS_higgswidth,CMS_higgswidth}. Both points are beyond the author's ability. 

\begin{table}
    \caption{\label{tab:higgswidth}
        Summary of Higgs mass and width measurments for different methods (unit: GeV). 
    }
    \begin{ruledtabular}
        \begin{tabular}{l | l l l}
            Parameter & Nominal (on-shell region) & F2F (on-shell region) & F2F (long region) \\
            \hline
            $m_H$ & $124.99\pm0.34$ & $125.05\pm0.45$ & $125.03\pm0.45$ \\
            $\Gamma_H$ (68.3\%) & 0.61 & 0.31 & 0.27 \\
            $\Gamma_H$ (90\%) &  1.54 &  1.04 & 0.92 \\ 
        \end{tabular}
    \end{ruledtabular}
\end{table}

\section{Application~2: new resonance search}\label{sec:newres}
In this section, we will use a simple example to illustrate how to use this method to search for new resonance. Suppose we are searching for a new particle, decaying to a pair of leptons, on the spectrum of the di-lepton mass, $m_{ll}$, from 100~GeV to 2~TeV. We further assume that the di-lepton mass resolution is $2~\%\times m_{ll}$.

\begin{figure}
    \includegraphics[width=0.45\textwidth]{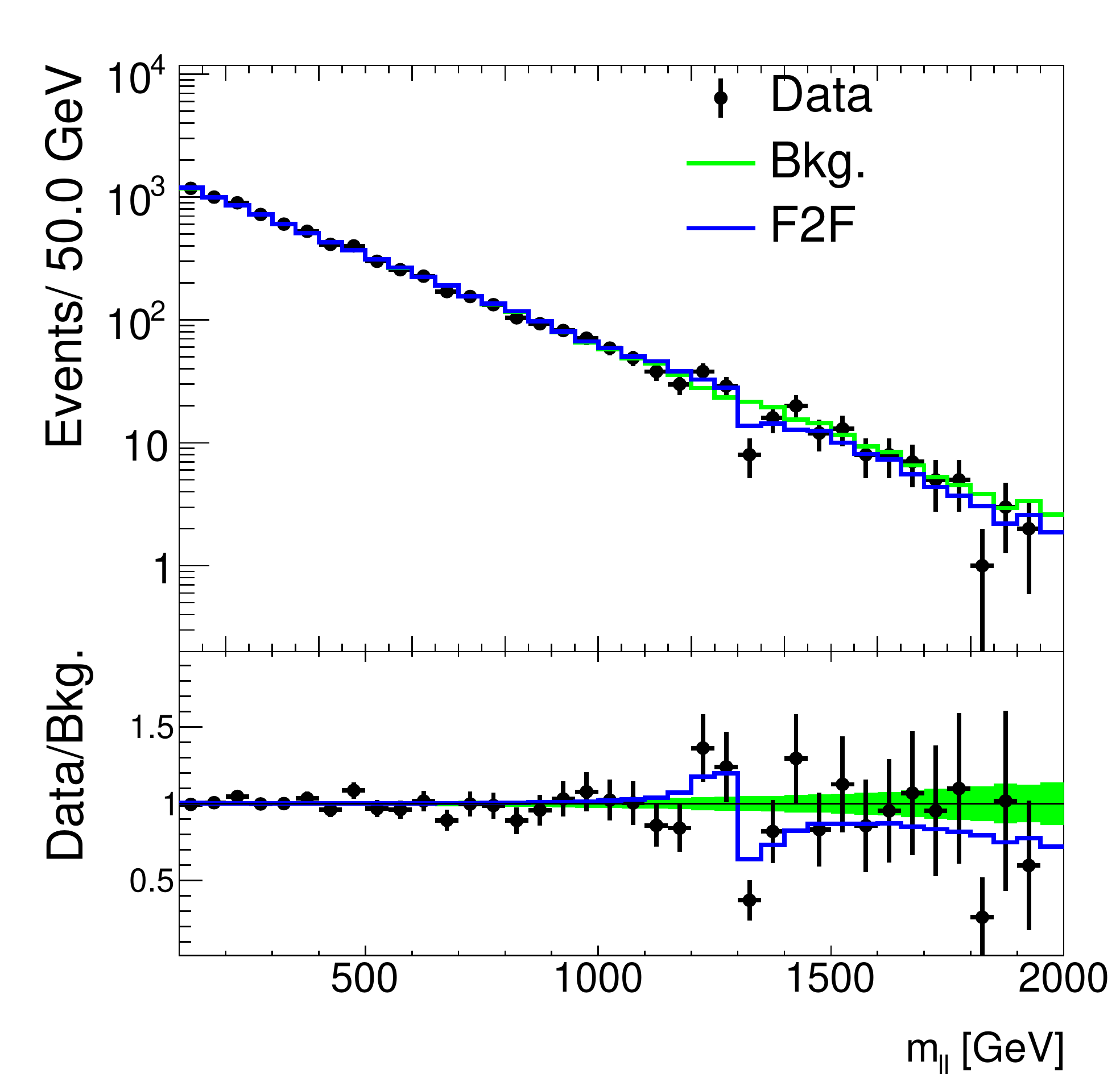}
    \includegraphics[width=0.45\textwidth]{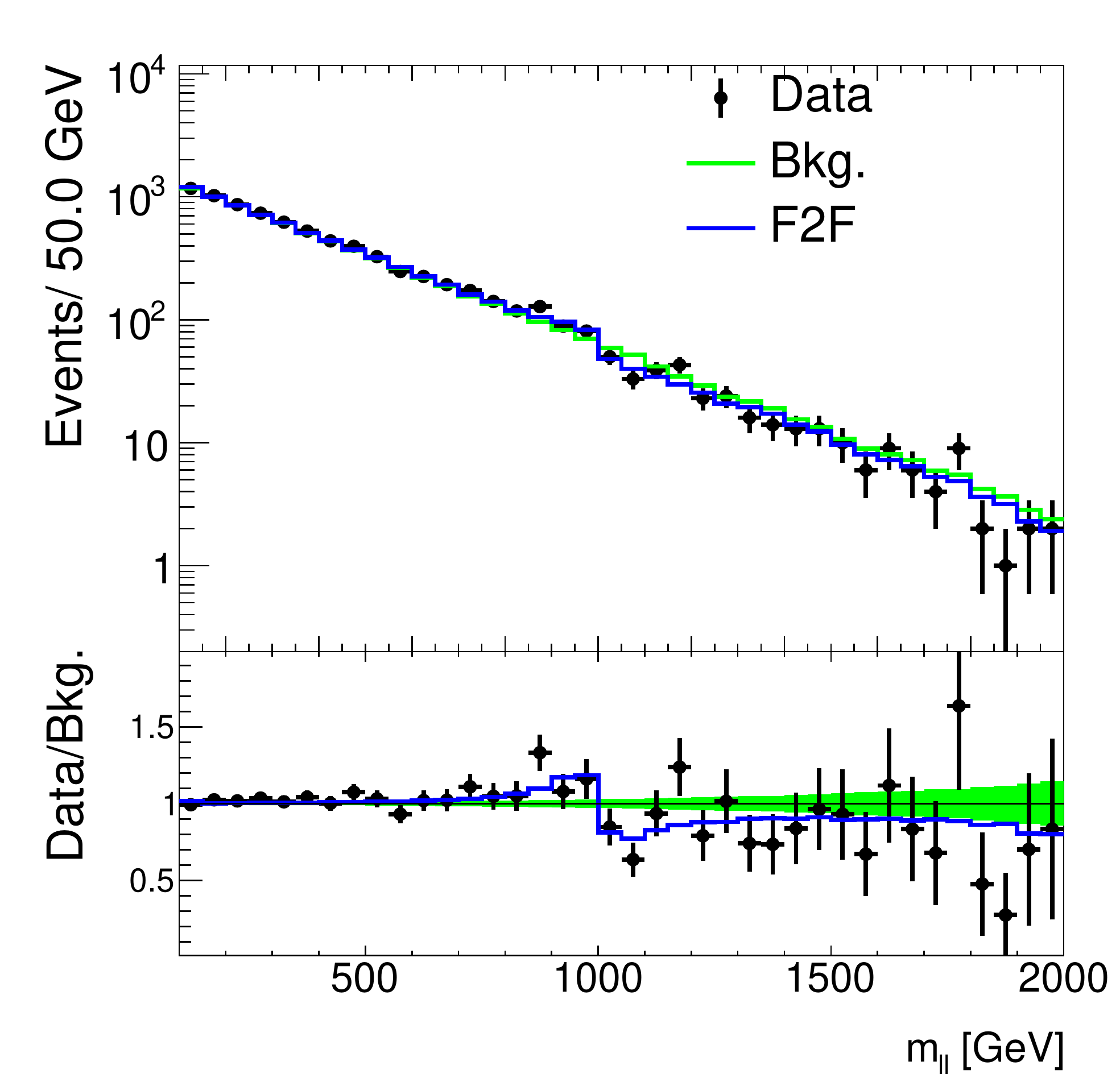}
    \caption{\label{fig:newres_fit}
        (color online)
        Fit to the toy data events generated under the background-only hypothesis (L) and background-plus-signal hypothesis (R). The left plot shows the fit with a resonance of mass 1.3~TeV and width 5~GeV. The right plot shows the fit with a resonance of mass 1~TeV and width 50~GeV. In the upper pads, the black dots with error bar represent data; the green histograms represent the background; the blue curves represent the best fit. The ratio between data and background is shown in the lower
        pads. The green band indicates the background MC uncertainty (the uncertainty itself is not considered in the fit). 
    }
\end{figure}

\begin{figure}
    \includegraphics[width=0.45\textwidth]{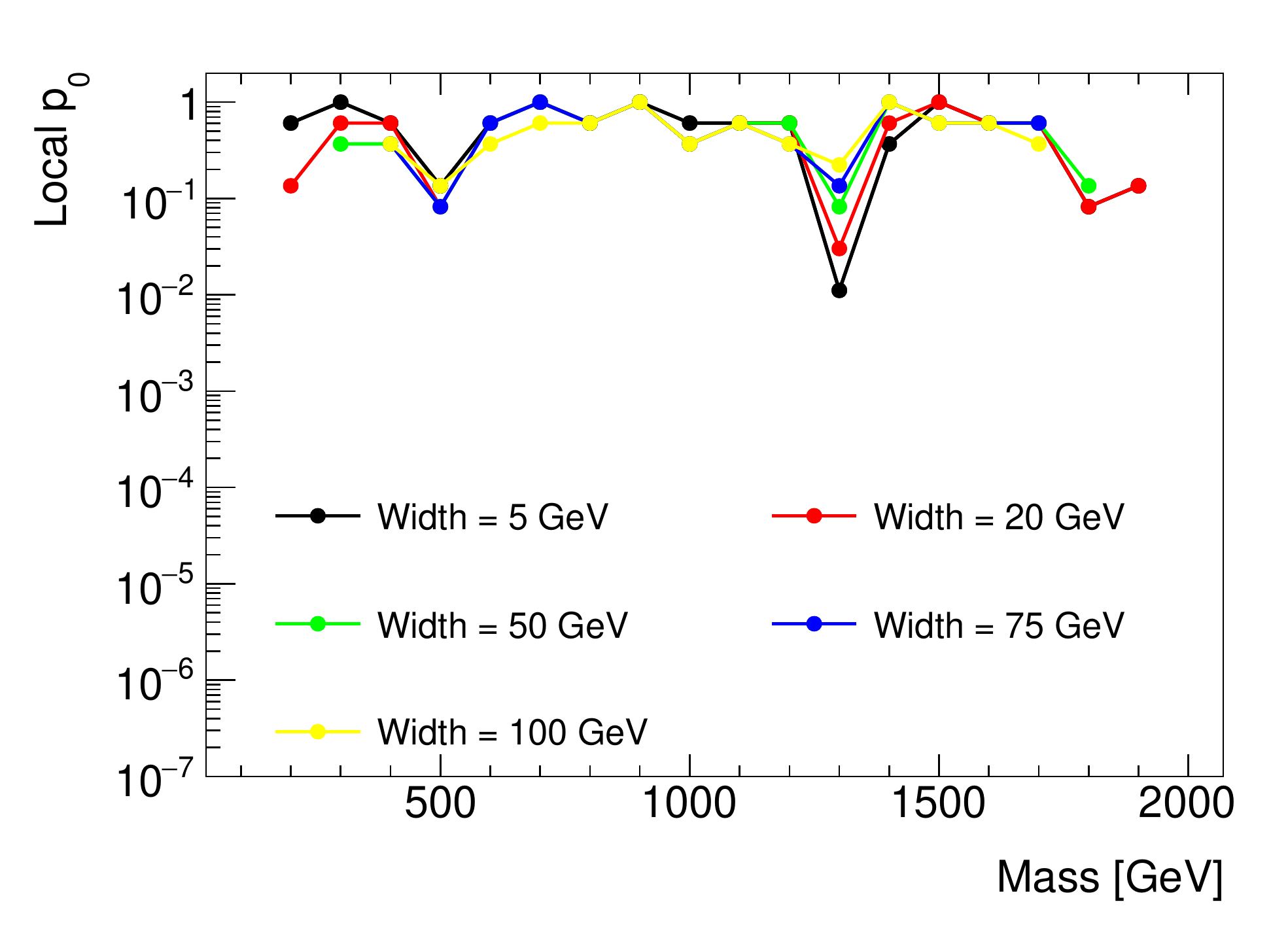}
    \includegraphics[width=0.45\textwidth]{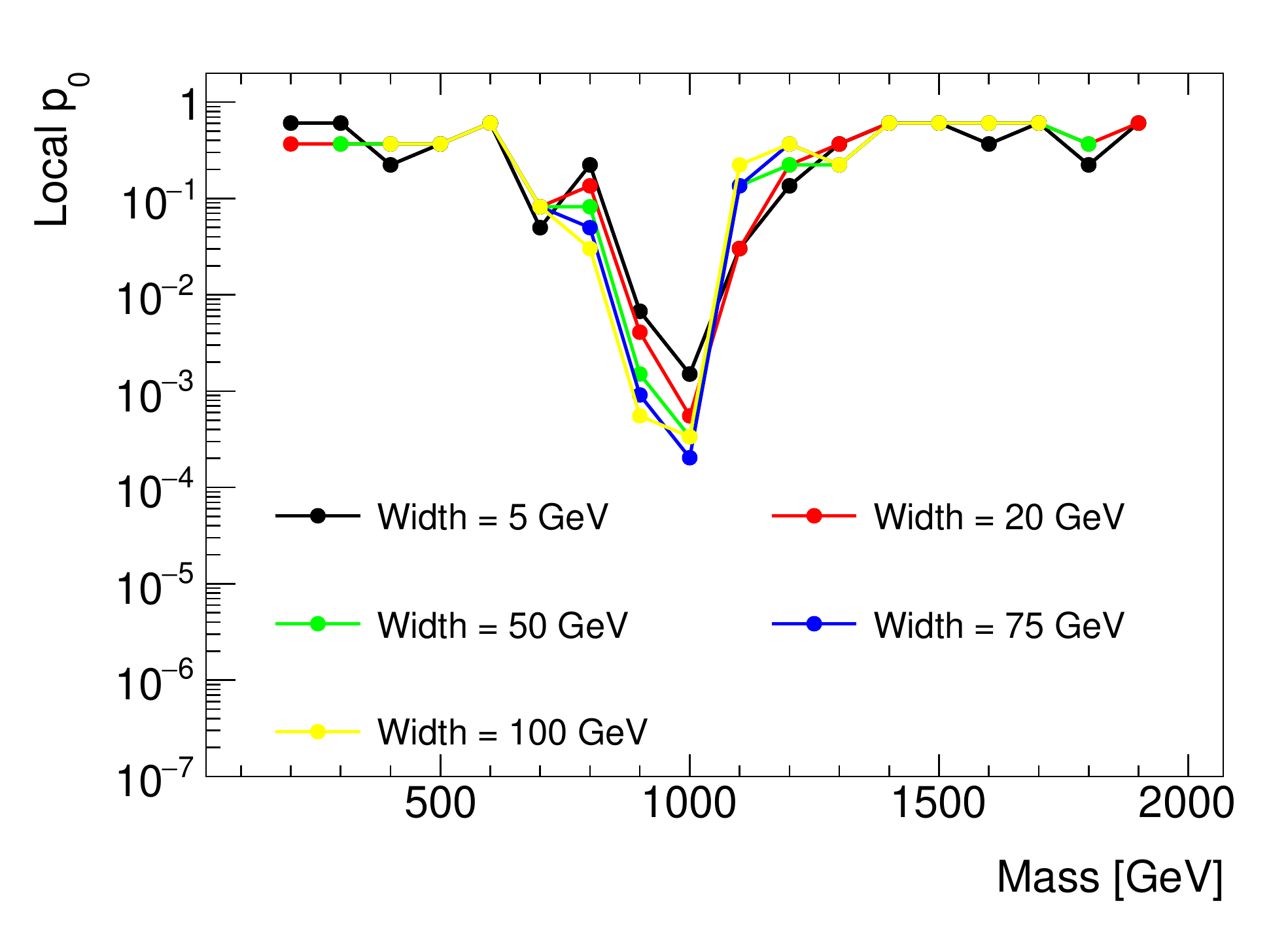}
    \caption{\label{fig:newres_p0}
        (color online)
        Local $p_0$ values for different masses and widths from fits to toy data events generated under background-only hypothesis (L) and background-plus-signal hypothesis (R).  
    }
\end{figure}

Two toy data samples are produced under the background-only hypothesis and the background-plus-signal hypothesis with inputing a resonance with mass 1~TeV and width 50~GeV, respectively. They are shown in Fig.~\ref{fig:newres_fit}. To hunt for a new particle with unknown mass and width, we perform multiple fits with mass changing from 200~GeV to 1.9~TeV with a step 100~GeV and width changing from 5~GeV to 100~GeV. Combining these fits and the background-only fit allows us to obtain the significance of a possible
resonance for given mass and width. This is the local $p_0$ value and shown in Fig.~\ref{fig:newres_p0}. The left plot shows the $p_0$ values for the background-only toy data sample. There seems to be a signal with mass 1.3~TeV and width 5~GeV but the significance is small. The corresponding fit is shown in the left plot in Fig.~\ref{fig:newres_fit}. The right plot of Fig.~\ref{fig:newres_p0} shows the $p_0$ values for the background-plus-signal toy data sample. A signal with mass 1~TeV is clearly seen. This is consistent with our input. The corresponding fit is shown in the right plot of Fig.~\ref{fig:newres_fit}.

To be exact, the calculation formula for the $p_0$ value is shown below.
\begin{equation}
    p_0(M,\Gamma) = \int_{\chi_b^2-\chi_{s+b}^2}^{+\infty}f_2(x) dx \:, 
\end{equation}
where $\chi_b^2$ ($\chi_{s+b}^2$) is the best-fit $\chi^2$ value defined in Eq.~\ref{eq:chi2} in the background-only (background-plus-signal) fit; $f_2(x)$ is the chi-square distribution function with 2 degrees of freedom (because we have 2 extra floating parameters after fixing $M$ and $\Gamma$ under the background-plus-signal hypothesis in Eq.~\ref{eq:fitfunc}).

\section{Summary}\label{sec:summary}
The interference effect is believed to be more and more important in future seaches for new particles. It is crucial to determine the mass and width of the new particle even if the interference makes the new particle show up with strange shape. In this work, a model-indepdent method, F2F, is proposed to search for an unknown particle and measure its mass and width when we do not have enough knowledge about the interference details. 
We express the sum of resonant signal and the interference as a cosine Fourier series and relate the Fourier coefficients with the mass and width. The detector resolution effect can be considered conveniently. No information on the signal model is needed. 
Toy experiments show good agreement between the results from F2F and the inputs. The precision is close to that from fitting using an explicit signal model. We also show that the method can be used to measure the SM Higgs width and to make statistic interpretation in searching for new resonance allowing for interference.

\section{Acknowledgement}
I would like to thank W.M. Song and N. Kauer for the instructions about the generator gg2VV. Special thanks goes to the HEP group in Warwick where I enjoy good time. This work will not come out without their support and the freedom atmosphere there. As usual, I would like to thank Fang Dai for encouraging words.


\begin{thebibliography}{99}
    \bibitem{phipi0}
        BESIII Collaboration, Phys. Rev. D91 (2015) 112001, arXiv:1504.03194
    \bibitem{ATLAS_hzz}
        ATLAS Collaboration, JHEP 1904 (2019) 048, arXiv:1902.05892
    \bibitem{ATLAS_vlq}
        ATLAS Collaboration, JHEP 1905 (2019) 164, arXiv:1812.07343
    \bibitem{CMS_zz}
        CMS Collaboration, JHEP 1806 (2018) 127, Erratum: JHEP 1903 (2019) 128, arXiv:1804.01939
    \bibitem{CMS_ttbar}
        CMS Collaboration, CERN-EP-2019-147, arXiv:1908.01115
    \bibitem{ggzz}
        S. Jung, Y.W. Yoon, and J. Song, Phys. Rev. D93 (2016) 055035, arXiv:1510.03450
    \bibitem{ttbar}
        A. Djouadi, J. Ellis, A. Popov, J. Quevillon, JHEP 1903 (2019) 119, arXiv:1901.03417
    \bibitem{song}
        N. Kauer, A. Lind, P. Maierh\"{o}fer, and W. Song, JHEP 1907 (2019) 108, arXiv:1905.03296
    \bibitem{loopint}
        see for example \url{https://en.wikipedia.org/wiki/Contour_integration}
    \bibitem{gg2VV0}
        T. Binoth, N. Kauer, and P. Mertsch, arXiv:0807.0024
    \bibitem{gg2VV1}
        N. Kauer and G. Passarino, JHEP 08 (2012) 116, arXiv:1206.4803
    \bibitem{madgraph}
        J. Alwell et al., JHEP 1407 (2014) 079, arXiv:1405.0301
    \bibitem{ATLAS_higgswidth}
        ATLAS Collaboration, Phys. Lett. B 786 (2018) 223, arXiv:1808.01191
    \bibitem{CMS_higgswidth}
        CMS Collaboration, Phys. Rev. D 99, 112003, arXiv:1901.00174   
    \bibitem{CMS_4muresolution}
        CMS Collaboration, JHEP 11 (2017) 047, arXiv:1706.09936, see additional figures in \url{http://cms-results.web.cern.ch/cms-results/public-results/publications/HIG-16-041}
\end{thebibliography}
\end{document}